\documentclass[a4paper]{article}

\usepackage[T1]{fontenc}
\usepackage{graphicx}
\usepackage{hyperref}
\usepackage{amsmath}
\usepackage[capitalize]{cleveref}
\usepackage{color}
\usepackage{fullpage}

\usepackage{optidef}
\usepackage{amssymb,amsthm}
\usepackage{enumerate} % allows \begin{enumerate}[(a)] style labels

\usepackage{mathtools}
\usepackage{tikz}
\usetikzlibrary{arrows.meta,positioning,calc,decorations.pathreplacing}

\newcommand{\I}{\mathcal{I}}
\newcommand{\cA}{\mathcal{A}}
\newcommand{\Cr}{\mathcal{M}}
\newcommand{\Cl}{\operatorname{Cl}}
\newcommand{\OPT}{\mathrm{OPT}}
\newcommand{\eps}{\varepsilon}
\newcommand{\val}{\operatorname{val}}
\newcommand{\one}{\mathbf{1}}
\newcommand{\MaxIMC}{\textsc{MaxIMC}}
\newcommand{\MaxEIMC}{\textsc{MaxEIMC}}
\definecolor{selblue}{HTML}{1F5A96}
\definecolor{famteal}{HTML}{197568}
\definecolor{famorange}{HTML}{B4581E}
\definecolor{fampurple}{HTML}{6B3FA0}

\theoremstyle{plain}
\newtheorem{theorem}{Theorem}
\newtheorem{lemma}{Lemma}
\newtheorem{claim}{Claim}
\Crefname{claim}{Claim}{Claims}

\theoremstyle{definition}
\newtheorem{definition}{Definition}

\theoremstyle{remark}
\newtheorem{remark}{Remark}

\usepackage{algorithm}
\usepackage[noend]{algpseudocode}
\usepackage{subcaption}

\usepackage{authblk}
\usepackage{orcidlink}

\title{NP-Hardness and a PTAS for Maximum Interval Multicover, with Applications to Dense Subgraphs}
\author[1]{Morteza Alimi \orcidlink{0000-0003-3732-2903}}
\author[1]{Tobias M\"{o}mke \orcidlink{0000-0002-2509-6972}}
\author[2]{Alexandru Popa \orcidlink{0000-0003-3364-1210}}
\author[1]{Michael Ruderer \orcidlink{0009-0009-1968-4821}}

\affil[1]{University of Augsburg}
\affil[2]{University of Bucharest}
\begin{document}
\maketitle   
\begin{abstract}
    Given intervals and points on a line, each point with a demand, and a budget $K$, Maximum Interval Multicover asks for at most $K$ intervals that satisfy as many points as possible, a point being satisfied if it lies in at least as many of the chosen intervals as its demand. This models, for instance, committee elections in which every candidate is approved by a contiguous group of voters and every voter wants at least her own number of approved candidates elected. Xu, Ran and Zhang introduced the problem and gave a $1/2$-approximation; the objective is not submodular when demands exceed one. Li, Tu, Zhang and Cao gave a polynomial-time approximation scheme (PTAS) for the minimization counterpart, unweighted partial interval multicover, and asked for its complexity. As the two problems share a decision version, the complexity of both was open. We prove that both problems are NP-hard, even for intervals of equal length; as their optima are integers bounded by the input size, neither has a fully polynomial-time approximation scheme unless $\mathrm{P}=\mathrm{NP}$. We then give a PTAS for Maximum Interval Multicover, also for arbitrary nonnegative nondecreasing rewards for the coverage of each point. Together, these results settle its approximability up to the existence of an efficient PTAS. Building on a decomposition from Nonner's PTAS for Densest $k$-Subgraph on interval graphs, our scheme cuts the line into pieces at a small loss and solves every piece exactly. Its dynamic program processes the endpoints of the intervals through a common point not along the line but in an interleaving of left and right endpoints that it chooses, which reduces its memory of these intervals to two counters. The same algorithm yields PTASes on convex bipartite graphs for Bipartite Dense Subgraph, improving on the ratio $2/3$ of Backer and Keil for the subclass of bipartite permutation graphs, and on interval graphs for Maximum $k$-Vertex Cover and weighted clique counts.
\end{abstract}

\clearpage
\section{Introduction}\label{sec:intro}
Consider a family of intervals and a set of points on the real line. 
Selecting a given number of intervals to cover as many points as possible is polynomial-time solvable~\cite{cohen-gonen2019,mettanant2026}.
We study the generalization in which each point has a demand specifying how many selected intervals must contain it.
Formally, we are given a family $\I$ of $n$ closed intervals and a set $P$ of $m$ points on the real line, each with a demand $q_p\in\mathbb{Z}_{\ge0}$.
We write $c_S(p)=|\{I\in S:p\in I\}|$ for the coverage of point $p \in P$ by an interval selection $S\subseteq\I$.
The point $p$ is \emph{satisfied} if $c_S(p)\ge q_p$.
In \emph{Maximum Interval Multicover} (\MaxIMC), introduced by Xu, Ran and Zhang~\cite{xrz-aaim,xrz-tcs}, the goal is to satisfy as many points as possible using at most $K$ intervals. 

A point contributes nothing until its whole demand is met, so intervals complement each other: e.g. two intervals that cover a point of demand 2 are worthless separately and valuable together.
Satisfying \emph{all} coverage demands with a minimum number of intervals is polynomial-time solvable.
Hence, the difficulty lies in choosing which demands to satisfy.

The model also describes committee elections in which every candidate is approved by a contiguous group of voters (the voter-interval domain~\cite{elkind-lackner2015}): electing $K$ candidates to maximize the number of voters who approve at least their own prescribed number of them is an instance of \MaxIMC. For satisfaction functions that are concave in the number of approved members, as in the Thiele rules studied there, polynomial-time algorithms on this domain were given in two recent preprints~\cite{manurangsi-sornat2026,avramidis2026}. \Cref{thm:ptas} below allows arbitrary nondecreasing satisfaction functions that may differ between voters.

Xu, Ran and Zhang gave a $1/2$-approximation for \MaxIMC{}~\cite{xrz-aaim,xrz-tcs}. Li, Tu, Zhang and Cao~\cite{li-tu-zhang-cao2026} studied its minimization counterpart,  Partial Interval Multicover, where the number of satisfied points is fixed and the number of selected intervals has to be minized. They gave a PTAS for it~\cite[Theorem~1]{li-tu-zhang-cao2026}, proved NP-completeness of a weighted generalization~\cite[Theorem~2]{li-tu-zhang-cao2026}, and stated that the complexity of the unweighted problem remained open. 
They also asked whether their approach could improve the approximation for \MaxIMC{}, and whether an approximation-preserving reduction relates partial interval multicover to Densest $k$-Subgraph. 
Their hardness proof also applies to a weighted version of Densest $k$-Subgraph. Their unweighted problem has the same decision version as \MaxIMC{}: can at most $K$ intervals satisfy at least $Q$ points? The variant of \MaxIMC{} in which a point counts only if it is covered \emph{exactly} as often as its demand is NP-hard already for unit rewards~\cite{cohen-gonen-levin-onn2017}.

We settle the complexity of \MaxIMC{} and give an approximation scheme that also applies to several dense subgraph problems.

\subsection{Our results}\label{sec:results}

We consider \MaxIMC{} with general coverage rewards: each point $p$ has an explicitly tabulated nondecreasing function $f_p:\{0,\dots,K\}\to\mathbb{Q}_{\ge0}$, and the objective is to maximize $F(S)=\sum_{p\in P}f_p(c_S(p))$ over all $S\subseteq\I$ with $|S|\le K$. Write $\OPT$ for this maximum.

\begin{theorem}\label{thm:ptas}
There is a PTAS for \MaxIMC{} with general coverage rewards.
\end{theorem}

With nonnegative point profits $w_p$ and $f_p(d)=w_p\cdot\one[d\ge q_p]$, \Cref{thm:ptas} gives a PTAS for \MaxIMC{}. This improves the ratio $1/2$ of~\cite{xrz-aaim,xrz-tcs} and positively answers the question of Li et al., albeit using a different method. 
Running their PTAS for every number $Q$ of points yields a selection that satisfies at least $\OPT$ many points but may use up to $(1+\eps)K$ intervals.
As points count only when fully covered, trimming such a solution to $K$ intervals may leave only one satisfied point wheres an optimal solution satisfies $\eps K+1$, even if all demands are at most $K$.(\Cref{app:trim}).

\begin{theorem}\label{thm:hardness}
The decision version of \MaxIMC{} is strongly NP-complete. This holds even when all intervals are distinct and have the same positive length, all endpoints and points are distinct integers, no point is an endpoint, and $1\le q_p\le\min\{K,c_{\I}(p)\}$ for every point $p$.
\end{theorem}

Zhou et al.~\cite{zhou-et-al2026} independently proved that this decision
problem is NP-complete even for unit intervals.
\Cref{thm:hardness} settles the complexity of \MaxIMC{} and, since the decision versions coincide, answers the question of Li et al.\ for unweighted partial interval multicover, strengthening their hardness result for the weighted version. In general instances, coordinates can be replaced by their ranks and demands and the budget capped at $n$, so all numbers are polynomially bounded anyway; ``strongly'' matters for intervals of equal length, whose coordinates cannot be replaced by ranks. As the optima are integers bounded by the input size, neither \MaxIMC{} nor partial interval multicover has an FPTAS unless $\mathrm{P}=\mathrm{NP}$; in particular, the PTAS of Li et al.\ cannot be improved to an FPTAS. Whether \MaxIMC{} has an efficient PTAS is open. With demands at most $q_{\max}$, a sweep solves \MaxIMC{} exactly with $(n+1)^{q_{\max}}\cdot\mathrm{poly}(n,m)$ operations (\Cref{thm:bounded}), so hardness requires unbounded demands unless $\mathrm{P}=\mathrm{NP}$.

The algorithm of \Cref{thm:ptas} extends to sums of nonnegative rewards of three kinds (\Cref{thm:general}): (i) for every point, a reward nondecreasing in its coverage; (ii) for every interval, a reward depending on the number of other selected intervals that contain its left endpoint and on whether the interval itself is selected, nondecreasing in both; (iii) for every \emph{optional point}, which can be selected under a separate or a joint budget, a reward nondecreasing in its coverage for either selection status. Kind~(ii) expresses edges and cliques of interval graphs, since two intervals intersect exactly when one contains the left endpoint of the other, and kind~(iii) the second side of a convex bipartite graph. The extension requires some optimal solution to be \emph{upward closed}, that is, to contain every interval that contains one of its members (with ties broken as in \Cref{sec:prelim}); for the objectives of \Cref{thm:ptas,thm:dense}, this follows from an exchange argument (\Cref{lem:exchange}). A bipartite graph $G=(U\cup V,E)$ is \emph{convex} if $V$ has a linear order in which the neighbourhood of every vertex of $U$ is consecutive. \emph{Bipartite Dense Subgraph} asks for $S_U\subseteq U$ and $S_V\subseteq V$ maximizing the number $|E(S_U,S_V)|$ of edges between them, subject either to separate budgets $|S_U|\le K_U$ and $|S_V|\le K_V$ or to a total budget $|S_U|+|S_V|\le K$. For a vertex set $S$ of a graph, $N_r(S)$ is the number of $r$-cliques induced by $S$, and for rationals $0=\mu_0\le\mu_1\le\mu_2$ the \emph{edge utility} of $S$ is $\sum_{\{u,v\}\in E}\mu_{|\{u,v\}\cap S|}$.

\begin{theorem}\label{thm:dense}
For every fixed $\eps\in(0,1)$, each of the following problems admits a deterministic algorithm that respects the budgets, achieves at least $(1-\eps)$ times the optimum, and uses $(N+1)^{O(1/\eps)}$ arithmetic operations on numbers of polynomial bit length, where $N$ is the number of vertices.
\begin{enumerate}[(a)]
\item Bipartite Dense Subgraph on convex bipartite graphs, with separate or total budgets.
\item On interval graphs, choosing at most $K$ vertices to maximize $\sum_{r=2}^{K}\lambda_rN_r(S)$, where the rationals $\lambda_r\ge0$ are part of the input.
\item On interval graphs, choosing at most $K$ vertices to maximize the edge utility, for given rationals $0\le\mu_1\le\mu_2$.
\end{enumerate}
\end{theorem}

For the total-budget version of part~(a), which is Densest $k$-Subgraph, Backer and Keil~\cite{backer-keil2010} gave a $3/2$-approximation, that is, value at least two thirds of the optimum, on bipartite permutation graphs, which form a proper subclass of the convex bipartite graphs~\cite{shrestha-tayu-ueno2011}; the same ratio for separate budgets was announced in a conference abstract by Inaba, Tayu and Ueno~\cite{inaba-tayu-ueno2016}. Backer and Keil noted that ``the ability to solve for a fixed but arbitrarily wide subproblem in polynomial time would lead to a PTAS''~\cite[Section~4]{backer-keil2010}; in this spirit, our window algorithm (\Cref{thm:window}) solves subproblems of any fixed width exactly, which yields a PTAS on the larger class. Parts~(b) and~(c) contain Densest $k$-Subgraph ($\lambda_2=1$ and $\lambda_r=0$ otherwise, or $\mu_1=0$ and $\mu_2=1$), for which Nonner gave a PTAS on interval graphs~\cite{nonner2011,nonner2016}; part~(c) also contains Maximum $k$-Vertex Cover, $k$ vertices covering the most edges ($\mu_1=\mu_2=1$), for which we know of no previous approximation scheme on interval graphs. Convex orders and interval representations can be computed in linear time~\cite{booth-lueker1976}. The complexity of Densest $k$-Subgraph on interval graphs is a long-standing open question~\cite{nonner2016}, and we do not know whether any problem of \Cref{thm:dense} is NP-hard; our framework does not give the reduction asked for by Li et al.

\subsection{Technical overview}\label{sec:overview}

\paragraph*{Slabs and shifting.}
We use the structure of an optimal solution from Nonner's PTAS for Densest $k$-Subgraph~\cite{nonner2016}. For the objective of \Cref{thm:ptas}, some optimal solution $O$ is upward closed, by the exchange argument of~\cite{nonner2016} and~\cite[Proposition~1]{li-tu-zhang-cao2026} (\Cref{lem:exchange}). Disjoint representatives, chosen greedily from $O$ by leftmost right endpoint, cut the line at their right endpoints into \emph{slabs}. An interval of $O$ covering a point of a slab contains one of the three nearest representatives, and is then forced by upward closure, or belongs to one of the two adjacent \emph{crossing families}: the intervals that start inside a representative and end between its right endpoint and that of the next. By shifting~\cite{hochbaum-maass1985,baker1994} with $L\approx1/\eps$, ignoring every $L$-th slab costs at most $\OPT/L$ for a suitable offset and leaves runs of fewer than $L$ slabs. The subproblem on a run, a \emph{window}, must be solved within its share of the budget and with little loss.

\paragraph*{Why crossing families are hard.}
A family of intervals that all cross a common point can be used in exponentially many ways, even with a fixed number of selected members and a \emph{closed} choice, which contains every member of the family that contains one of its members: if the starts and the ends appear in the same order, every subset is closed, and subsets of the same size produce different coverage patterns. Neither Nonner's scheme nor that of Li et al.\ optimizes over this space exactly. Nonner keeps a polynomial collection of choices of the prescribed size whose coverage is close to that of the optimal choice; a small loss of coverage costs few edges, but it can destroy a threshold reward completely. Li et al.\ replace the optimal choice by one that dominates its coverage at every point, at the price of up to $(1+\eps)$ times as many intervals outside the forced containers~\cite[Lemma~1]{li-tu-zhang-cao2026}; this is affordable when the number of intervals is the objective, but not under a hard budget. A dynamic program that sweeps the line from left to right would have to remember the selected crossing intervals.

\begin{figure}[t]
\centering
\begin{tikzpicture}[x=0.9cm,y=0.4cm,font=\small]
  \draw[->,gray] (0,0)--(6.2,0);
  \draw[line width=2.4pt] (0.3,4.4)--(3.0,4.4) node[midway,above] {$A$};
  \draw[dashed,gray] (3.0,-0.4)--(3.0,5.1) node[above] {$r(A)$};
  \foreach \s/\e/\y/\i/\sel in {0.8/4.2/0.8/1/1, 1.35/3.6/1.6/2/1, 1.9/5.4/2.4/3/1, 2.45/4.8/3.2/4/0}{
    \ifnum\sel=1
      \draw[selblue,line width=1.8pt] (\s,\y)--(\e,\y);
    \else
      \draw[gray,line width=1.1pt,dash pattern=on 3pt off 2pt] (\s,\y)--(\e,\y);
    \fi
    \node[left,font=\scriptsize] at (\s,\y) {$x_{\i}$};
  }
  \node[font=\scriptsize,anchor=north] at (1.6,-0.1) {starts};
  \node[font=\scriptsize,anchor=north] at (4.5,-0.1) {ends};
  \node[font=\bfseries\small] at (0.0,5.2) {(a)};
  \begin{scope}[xshift=8.6cm,yshift=0.0cm,x=0.5cm,y=0.5cm]
    \foreach \i in {0,...,4}{\draw[gray!35] (\i,0)--(\i,4); \draw[gray!35] (0,\i)--(4,\i);}
    \foreach \j in {1,...,4}{\node[font=\scriptsize] at (\j-0.5,4.35) {$t_{\j}$};}
    \foreach \i in {1,...,4}{\node[font=\scriptsize] at (-0.4,4.5-\i) {$s_{\i}$};}
    \foreach \i/\j/\sel in {1/2/1,2/1/1,3/4/1,4/3/0}{
      \ifnum\sel=1 \fill[selblue] (\j-0.5,4.5-\i) circle (4.5pt);
        \node[font=\tiny,text=white,inner sep=0pt] at (\j-0.5,4.5-\i) {$\i$};
      \else \draw[gray,thick] (\j-0.5,4.5-\i) circle (4.5pt);
        \node[font=\tiny,text=gray,inner sep=0pt] at (\j-0.5,4.5-\i) {$\i$}; \fi}
    \draw[line width=1.6pt,famorange,rounded corners=1pt] (0,4)--(0,2)--(2,2)--(2,1)--(3,1)--(3,0)--(4,0);
    \node[font=\bfseries\small] at (-1.3,5.0) {(b)};
    \node[font=\scriptsize,align=left,anchor=west] at (4.4,2.2) {merge:\\ $s_1\,s_2\,t_1\,t_2\,s_3\,t_3\,s_4\,t_4$};
  \end{scope}
\end{tikzpicture}
\caption{(a) A crossing family $x_1,\dots,x_4$ of intervals that start inside $A$ and end to the right of $r(A)$; $x_1$ contains $x_2$, and $x_3$ contains $x_4$. The selected members $x_1,x_2,x_3$ are solid. (b) Rows are the starts $s_1,s_2,\dots$ in increasing order, columns the ends; member $x_i$ (labelled $i$) sits in the row of its start and the column of its end; its containers lie above and to the right. The merge $s_1\,s_2\,t_1\,t_2\,s_3\,t_3\,s_4\,t_4$ is the monotone path (down for a start, right for an end); it selects the members whose start precedes their end, the cells to the right of the path, so it cannot select $x_2$ without $x_1$.}
\label{fig:merge}
\end{figure}
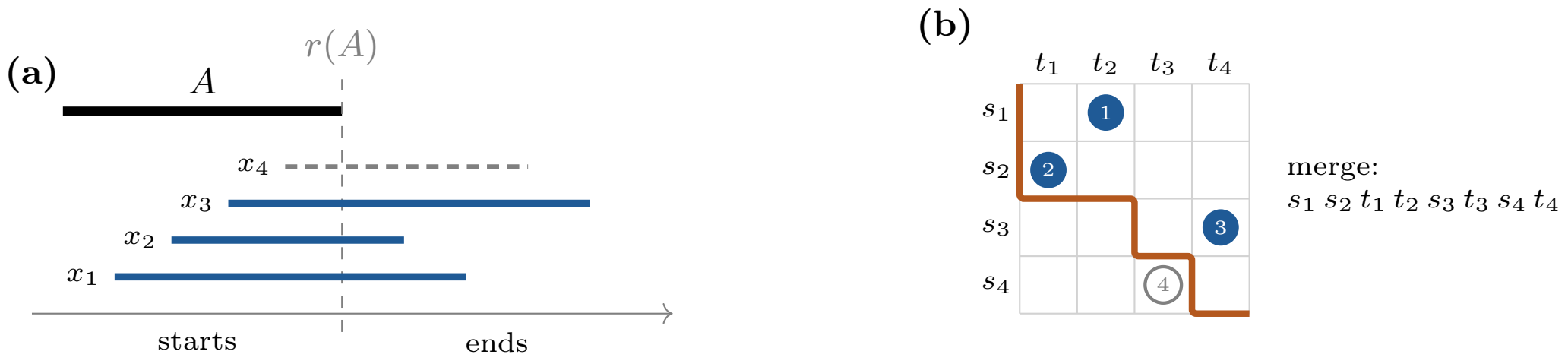

\paragraph*{Choices as merges.}
Sort the starts of the members of a crossing family and, separately, their ends. A \emph{merge} of the two sorted lists, that is, any interleaving of them, selects the members whose start precedes their end in it; the sets selected in this way are exactly the closed choices (\Cref{lem:merge}, \Cref{fig:merge}). This elementary fact is a form of the description of the up-sets of a two-dimensional order by monotone lattice paths. Our key observation is that a merge can serve as the order in which a dynamic program visits the endpoints. Unlike their order on the line, where all starts precede all ends, a merge advances on both sides of the common point at once, and the program need not store the choice: a point left of the common point is covered by the selected members whose starts have been visited, and a point to the right by those whose ends have not, whose number is the prescribed size minus the number of selected ends visited; two counters suffice.

\paragraph*{Chains of families.}
Consecutive crossing families of a window share a slab, which contains the ends of one family and the starts of the next. Merges chosen independently for all families can always be realized by one interleaving of the sorted event lists of all slabs (\Cref{lem:compatible}). Hence a dynamic program whose state records how many events of each slab have been processed, and two counters per family, optimizes exactly over all closed choices with prescribed family sizes, with $(n+m+1)^{O(L)}$ operations (\Cref{thm:window}). To combine windows, each is charged only for its intervals that do not contain its leftmost boundary, similar to the accounting of shared containers in Nonner's PTAS~\cite{nonner2016}; this charge bounds the number of intervals that the window adds to a solution assembled from left to right, and the charged sets of the windows of an optimal solution are disjoint (\Cref{lem:charges}). So a longest-path computation stays within the budget.

\paragraph*{Hardness.}
We reduce from rectilinear Steiner arborescence, which is strongly NP-complete~\cite{shi-su2006} (see also~\cite[Section~2]{aichholzer-mulzer-pilz2015}), via integral flows on a grid. We first build a \emph{proper} family of intervals, in which no interval contains another, with tests (guards and probes, \Cref{fig:packet}) that reward exact coverage values: packets of interchangeable candidate intervals carry flow values, and the tests copy and conserve flow and reward arcs without flow. Properness is what the next step needs: positive integer masses at positions on the line, at least one between any two consecutive endpoints, then give every interval the same total mass, and adding at every position a reward proportional to mass times coverage makes the tests nondecreasing while adding the same constant to every selection of $K$ intervals. Nondecreasing integer rewards are sums of threshold rewards, that is, points with demands, and reading the polynomially bounded masses as lengths makes all intervals equally long.

\subsection{Related work}\label{sec:related}

On general set systems, \MaxIMC{} with all demands equal to two contains Densest $k$-Subgraph (the sets are the vertices, the elements the edges), which is hard to approximate within almost polynomial factors under the Exponential Time Hypothesis~\cite{manurangsi2017}; much better ratios are achievable on structured classes~\cite{backer-keil2010,nonner2016}. If every element instead earns its coverage capped at a common value $q$, the objective is submodular, and a randomized algorithm achieves the ratio $1-q^q\mathrm{e}^{-q}/q!$, which is best possible under the Unique Games Conjecture~\cite{barman-et-al2022}. Maximum $k$-Vertex Cover has a rich approximation literature~\cite{ageev-sviridenko2004,manurangsi2019}, including a PTAS for bipartite graphs announced in a preprint~\cite{paschos2019}; solving it exactly on a graph with $N$ vertices is equivalent to solving Sparsest $(N-k)$-Subgraph exactly, whose complexity on interval graphs is open~\cite{watrigant-bougeret-giroudeau2016}. \Cref{sec:prelim,sec:window,sec:ptas} prove \Cref{thm:ptas}, \Cref{sec:general} treats general rewards and \Cref{thm:dense}, and \Cref{sec:hardness} proves \Cref{thm:hardness}; the appendix contains the omitted proofs.
\section{Preliminaries and the slab structure}\label{sec:prelim}

We fix conventions and describe the structure of an upward-closed optimal solution, adapting the decomposition of Nonner~\cite{nonner2016}.

\paragraph*{Conventions.}
Intervals are closed, coordinates are rational, and only their order matters. We refine ties once and for all: at a common coordinate, starts come first, then points (including the optional points of \Cref{sec:general}), then ends, and remaining ties are broken by fixed labels. Afterwards all endpoints and points are distinct, a point $p$ lies in an interval $I$ exactly when $\ell(I)<p<r(I)$, where $\ell(I)$ and $r(I)$ denote the endpoints, two intervals intersect exactly when they did before, and an interval that contains another after the refinement also contained it before. From now on, containment refers to the refined coordinates; repeated intervals in the input are therefore allowed. We assume $K\le n$, since otherwise selecting all intervals is optimal. For $S\subseteq\I$ and any position $x$ on the line, let
\[
c_S(x)\;=\;\bigl|\{I\in S:\ \ell(I)<x<r(I)\}\bigr|.
\]
At a point this is its coverage. At the start of an interval $J$ it counts the selected intervals that start before $J$ and are still active, excluding $J$ itself; this case is used only in \Cref{sec:general}. For intervals $A$ and $B$ we write $A\prec B$ if $r(A)<\ell(B)$. The \emph{containers} of an interval $A$, which need not belong to $\I$, form the set $\Cl(A)=\{J\in\I: J\supseteq A\}$. A set $S\subseteq\I$ is \emph{upward closed} if it contains all containers of each of its members. We write $f(S)=\sum_{p\in P}f_p(c_S(p))$ for the objective of \Cref{thm:ptas} and $\OPT$ for its maximum over $|S|\le K$. We extend every $f_p$ to $\{0,\dots,n\}$ by $f_p(d)=f_p(K)$ for $d>K$; this changes no value of a feasible solution.

\begin{lemma}[containment exchange~\cite{nonner2016},~{\cite[Proposition~1]{li-tu-zhang-cao2026}}]\label{lem:exchange}
Suppose that replacing a selected interval by an unselected container preserves feasibility and does not decrease the objective. Then some optimal solution is upward closed. In particular, this holds for $f$ and $|S|\le K$.
\end{lemma}

\begin{proof}
Number the intervals so that containers come after the intervals they contain. An optimal solution maximizing the sum of the numbers of its members is upward closed, since an exchange would increase this sum. For $f$, an exchange keeps $|S|$ and does not decrease any coverage.
\end{proof}

\paragraph*{Representatives and slabs.}
We add three \emph{sentinel} intervals $\bot\prec\top_1\prec\top_2$ such that $\bot$ lies to the left of all coordinates of the instance and $\top_1,\top_2$ lie to the right of them. They contain no point, are contained in no interval of $\I$, and are never selected; they only serve as boundaries. Let $\cA=\I\cup\{\bot,\top_1,\top_2\}$.

Fix an upward-closed optimal solution $O$. Its \emph{representatives} are chosen greedily: $I_1$ is the interval of $O$ with the smallest right endpoint, and $I_{i+1}$ is the interval of $O$ with the smallest right endpoint among those that start to the right of $r(I_i)$; the procedure stops when no such interval exists, after $h\ge0$ steps. The representatives are pairwise disjoint, $I_1\prec\dots\prec I_h$. We set $I_0=\bot$, $I_{h+1}=\top_1$ and $I_{h+2}=\top_2$. The \emph{slabs} are the open regions $\Sigma_i=(r(I_{i-1}),r(I_i))$ for $1\le i\le h+1$; every point lies in exactly one slab. For $A\prec B$, the \emph{crossing family} of $A$ and $B$ is
\[
\Cr(A,B)\;=\;\bigl\{J\in\I:\ \ell(A)<\ell(J)<r(A)<r(J)<r(B)\bigr\},
\]
the set of intervals that start inside $A$ and end between the right endpoints of $A$ and $B$. In particular $\Cr(\bot,B)=\emptyset$ for every $B$.

The next lemma says that few groups of intervals matter in a slab.

\begin{lemma}[locality]\label{lem:locality}
Let $1\le i\le h+1$, let $x$ be a position in $\Sigma_i$, and let $J\in O$ satisfy $\ell(J)\le x\le r(J)$. Then
\[
J\in \Cl(I_{i-1})\cup\Cl(I_i)\cup\Cl(I_{i+1})\cup\Cr(I_{i-1},I_i)\cup\Cr(I_i,I_{i+1}).
\]
\end{lemma}

\begin{proof}[Proof of \Cref{lem:locality}]
If $J$ is one of $I_{i-1},I_i,I_{i+1}$, it is its own container, so assume otherwise. We have $r(J)\ge x>r(I_{i-1})$, and $r(J)\ne r(I_i)$ because endpoints are distinct.

\emph{Case $r(J)<r(I_i)$.} If $J$ started to the right of $r(I_{i-1})$, then for $i\le h$ it would contradict the choice of $I_i$, which has the smallest right endpoint among the intervals of $O$ starting there, and for $i=h+1$ it would contradict the termination of the greedy procedure. Hence $\ell(J)<r(I_{i-1})$; in particular $I_{i-1}$ is not the sentinel $\bot$. If $\ell(J)<\ell(I_{i-1})$, then $J$ contains $I_{i-1}$. Otherwise $\ell(I_{i-1})<\ell(J)<r(I_{i-1})<r(J)<r(I_i)$, that is, $J\in\Cr(I_{i-1},I_i)$.

\emph{Case $r(J)>r(I_i)$.} Then $i\le h$, since no interval of $\I$ ends to the right of $r(\top_1)$, and $\ell(J)\le x<r(I_i)$. If $\ell(J)<\ell(I_i)$, then $J$ contains $I_i$. Otherwise $\ell(I_i)<\ell(J)<r(I_i)<r(J)$. If $r(J)<r(I_{i+1})$, then $J\in\Cr(I_i,I_{i+1})$. If $r(J)>r(I_{i+1})$, then $I_{i+1}$ is not a sentinel, and since $\ell(J)<r(I_i)<\ell(I_{i+1})$, the interval $J$ contains $I_{i+1}$.
\end{proof}

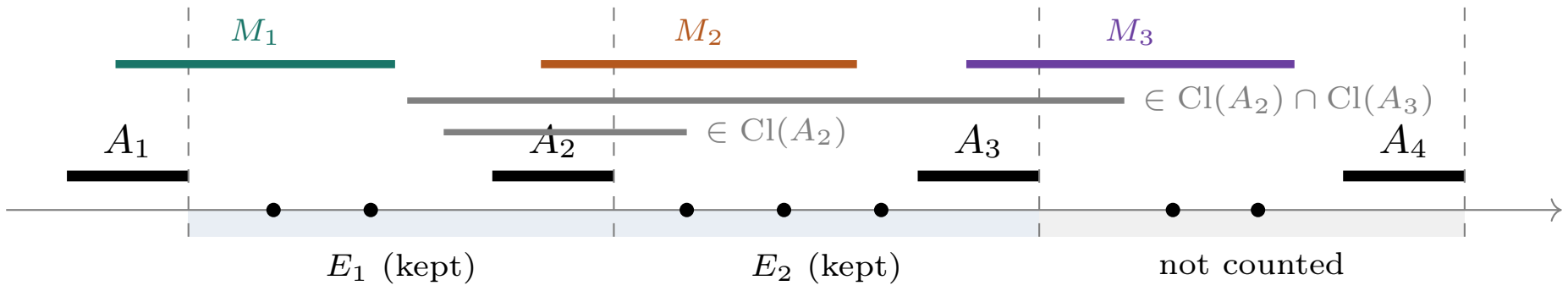
\begin{figure}[t]
\centering
\begin{tikzpicture}[x=0.95cm,y=0.38cm,font=\small]
  \fill[selblue!10] (1.5,-0.55) rectangle (8.5,0);
  \fill[gray!12] (8.5,-0.55) rectangle (12,0);
  \draw[->,gray] (0,0)--(12.8,0);
  \foreach \a/\b/\lab in {0.5/1.5/{A_1},4/5/{A_2},7.5/8.5/{A_3},11/12/{A_4}}{
    \draw[line width=2.4pt] (\a,0.7)--(\b,0.7);
    \node[above] at ({(\a+\b)/2},0.75) {$\lab$};
    \draw[dashed,gray] (\b,-0.55)--(\b,4.3);}
  % containers (fixed set)
  \draw[gray,line width=1.4pt] (3.6,1.6)--(5.6,1.6) node[right,font=\scriptsize] {$\in\Cl(A_2)$};
  \draw[gray,line width=1.4pt] (3.3,2.25)--(9.2,2.25) node[right,font=\scriptsize] {$\in\Cl(A_2)\cap\Cl(A_3)$};
  % one member of each family
  \draw[famteal,line width=1.8pt] (0.9,3.0)--(3.2,3.0) node[midway,above,font=\scriptsize] {$M_1$};
  \draw[famorange,line width=1.8pt] (4.4,3.0)--(7.0,3.0) node[midway,above,font=\scriptsize] {$M_2$};
  \draw[fampurple,line width=1.8pt] (7.9,3.0)--(10.6,3.0) node[midway,above,font=\scriptsize] {$M_3$};
  % points
  \foreach \x in {2.2,3.0,5.6,6.4,7.2,9.6,10.3}{\fill (\x,0) circle (1.6pt);}
  \node[below,font=\scriptsize] at (3.25,-0.55) {$E_1$ (kept)};
  \node[below,font=\scriptsize] at (6.75,-0.55) {$E_2$ (kept)};
  \node[below,font=\scriptsize] at (10.25,-0.55) {not counted};
\end{tikzpicture}
\caption{A window with $R=2$ kept slabs; the boundaries $A_1\prec\dots\prec A_4$ play the role of consecutive representatives. One member of each family $M_j=\Cr(A_j,A_{j+1})$ is shown, starting inside $A_j$ and ending in the next slab; the containers of the boundaries (gray) form the fixed set $F$. Within a window selection, a point of $E_t$ is covered only by intervals of $F$ and members of $M_t$ and $M_{t+1}$.}
\label{fig:window}
\end{figure}
\section{Exact optimization in a window}\label{sec:window}

This section contains the main technical contribution, an exact algorithm for windows. Their boundaries are arbitrary intervals; \Cref{sec:ptas} uses boundaries taken from an optimal solution, which the algorithm need not know.

\begin{definition}[window]\label{def:window}
Let $R\ge1$. A \emph{window} consists of intervals $A_1\prec A_2\prec\dots\prec A_{R+2}$ from $\cA$ and integers $a_1,\dots,a_{R+1}\ge0$. Its \emph{kept slabs} are $E_t=(r(A_t),r(A_{t+1}))$ for $1\le t\le R$, its \emph{families} are $M_j=\Cr(A_j,A_{j+1})$ for $1\le j\le R+1$, and its \emph{fixed set} is $F=\Cl(A_1)\cup\dots\cup\Cl(A_{R+2})$. A \emph{choice} consists of sets $Y_j\subseteq M_j$ with $|Y_j|=a_j$ for $1\le j\le R+1$; its \emph{window selection} is $W=F\cup Y_1\cup\dots\cup Y_{R+1}$, and its value is
\[
\val(W)\;=\;\sum_{p\in P\cap(E_1\cup\dots\cup E_R)} f_p\bigl(c_W(p)\bigr).
\]
\end{definition}

The intervals of $F$ are forced; the only freedom lies in the families (\Cref{fig:window}).

\begin{lemma}[structure of a window]\label{lem:structure}
In every window the following hold.
\begin{enumerate}[(i)]
\item The families are pairwise disjoint and disjoint from $F$. Hence the \emph{cost} $\kappa=|W\setminus\Cl(A_1)|=|F\setminus\Cl(A_1)|+\sum_{j}a_j$ is the same for all choices.
\item A member of $M_j$ starts inside $A_j$, hence in $E_{j-1}$ if $j\ge2$, and it ends in $E_j$ if $j\le R$, or between $r(A_{R+1})$ and $r(A_{R+2})$ if $j=R+1$.
\item For a position $x\in E_t$,
\(
c_W(x)=c_F(x)+\bigl|\{J\in Y_t:\ r(J)>x\}\bigr|+\bigl|\{J\in Y_{t+1}:\ \ell(J)<x\}\bigr|.
\)
\end{enumerate}
\end{lemma}

\begin{proof}[Proof of \Cref{lem:structure}]
A member $J$ of $M_j$ satisfies $\ell(A_j)<\ell(J)<r(A_j)<r(J)<r(A_{j+1})$. Hence it contains no boundary, and the range of $r(J)$ determines $j$, which gives~(i); (ii) follows since $r(A_{j-1})<\ell(A_j)$. For~(iii), members of $M_j$ with $j<t$ end before $r(A_t)<x$, members with $j>t+1$ start after $r(A_{t+1})>x$, members of $M_t$ start before $x$ and members of $M_{t+1}$ end after $x$.
\end{proof}

\begin{theorem}[exact window algorithm]\label{thm:window}
For every window with $a_j\le|M_j|$ for all $j$, a choice of maximum value, and its value, can be computed with $(n+m+1)^{O(R)}$ operations.
\end{theorem}

\subsection{Closed choices are merges}\label{sec:merges}

Fix one family $M=M_j$. All members contain $r(A_j)$, so a member $x'$ contains a member $x$ exactly when $\ell(x')<\ell(x)$ and $r(x)<r(x')$. A set $Y\subseteq M$ is \emph{closed} if it contains every member of $M$ containing a member of $Y$.

\begin{lemma}\label{lem:closed-suffice}
Every choice can be replaced by a choice with the same family sizes, in which every $Y_j$ is closed, whose window selection has at least the same coverage at every position. In particular, some choice of maximum value consists of closed sets.
\end{lemma}

\begin{proof}
Replace a selected $x\in Y_j$ by an unselected $x'\in M_j$ that contains it; this keeps $|Y_j|$ and does not decrease any coverage. As in the proof of \Cref{lem:exchange}, finitely many replacements lead to closed sets.
\end{proof}

A \emph{merge} of $M$ is a sequence of the $2|M|$ endpoints of its members that lists the starts in increasing order and the ends in increasing order. The following correspondence is elementary, a variant of the bijection between up-sets of a two-dimensional order and monotone lattice paths; what is new is its use as a dynamic-programming state. The merge \emph{selects} the members whose start precedes their end in it.

\begin{lemma}[merges]\label{lem:merge}
A set $Y\subseteq M$ is closed if and only if it is selected by some merge of $M$.
\end{lemma}

\begin{proof}[Proof of \Cref{lem:merge}]
For $x\in M$, let $\rho(x)$ and $\gamma(x)$ be the ranks of its start among the starts and of its end among the ends of $M$. A member $x'\ne x$ contains $x$ exactly when $\rho(x')<\rho(x)$ and $\gamma(x')>\gamma(x)$.

Suppose a merge selects $x$, and $x'$ contains $x$. In the merge, the start of $x'$ comes no later than the start of $x$, which precedes the end of $x$, which comes no later than the end of $x'$. So the merge selects $x'$, and the selected set is closed.

Conversely, let $Y$ be closed. For $1\le i\le|M|$ let $\theta_i=1+\max\{\gamma(x): x\in M\setminus Y,\ \rho(x)\le i\}$, where the maximum of the empty set is $0$; the sequence $\theta$ is nondecreasing. We claim that $x\in Y$ exactly when $\gamma(x)\ge\theta_{\rho(x)}$. If $x\notin Y$, then $x$ itself shows $\theta_{\rho(x)}\ge\gamma(x)+1$. If $x\in Y$, every $x'\notin Y$ with $\rho(x')\le\rho(x)$ satisfies $\gamma(x')<\gamma(x)$, since otherwise $x'$ would contain $x$ and closedness would force $x'\in Y$; hence $\theta_{\rho(x)}\le\gamma(x)$. Now build the merge that places the $i$-th start immediately after the first $\theta_i-1$ ends, keeping the starts in increasing order, which is possible because $\theta$ is nondecreasing. The start of $x$ then precedes its end exactly when $\theta_{\rho(x)}-1<\gamma(x)$, so the merge selects $Y$.
\end{proof}

\subsection{Interleaving all families}\label{sec:interleave}

The families of a window share slabs: by \Cref{lem:structure}(ii), the slab $E_t$ contains the ends of $M_t$ and the starts of $M_{t+1}$. We record the events of the window in $R+2$ \emph{event sequences}:
$\sigma_0$ lists the starts of $M_1$ in increasing order;
for $1\le t\le R$, $\sigma_t$ lists, in increasing order, the ends of $M_t$, the starts of $M_{t+1}$ and the points of $P$ in $E_t$;
and $\sigma_{R+1}$ lists the ends of $M_{R+1}$ in increasing order.
The start of a member of $M_j$ lies in $\sigma_{j-1}$ and its end in $\sigma_j$. An \emph{interleaving} is a linear order of all events that restricts to the order of each event sequence. Restricted to the endpoints of $M_j$, an interleaving is a merge of $M_j$; we say that the interleaving \emph{selects} the members selected by this merge.

\begin{lemma}[compatibility]\label{lem:compatible}
For any merges $\pi_1,\dots,\pi_{R+1}$ of $M_1,\dots,M_{R+1}$ there is an interleaving whose restriction to the endpoints of $M_j$ is $\pi_j$ for every $j$.
\end{lemma}

\begin{proof}
We construct linear orders $G_0,\dots,G_{R+1}$, where $G_t$ orders the events of $\sigma_0,\dots,\sigma_t$, restricts to the order of each of these sequences, and restricts to $\pi_j$ on the endpoints of $M_j$ for every $j\le t$. Let $G_0=\sigma_0$. Given $G_{t-1}$, let $\xi_1,\dots,\xi_b$ be the starts of $M_t$ in increasing order; they occur in this order in $G_{t-1}$, because they belong to $\sigma_{t-1}$. For each end of $M_t$, the merge $\pi_t$ prescribes how many of these starts precede it, and these numbers are nondecreasing along $\sigma_t$, because the ends appear in increasing order in $\pi_t$. Insert the events of $\sigma_t$ one by one, in their order, into the gaps between consecutive events of $G_{t-1}$, always into the gap marked by a cursor; events placed in the same gap keep the order of insertion. The cursor starts at the gap in front of all events of $G_{t-1}$. Before an end that must be preceded by exactly $q$ starts is placed, the cursor moves to the gap immediately after $\xi_q$ if it is not there already ($q=0$ leaves it at the front); every other event is placed at the current cursor. Because the prescribed numbers are nondecreasing, the cursor never moves to the left, so $\sigma_t$ keeps its order. Each end of $M_t$ is preceded by exactly the prescribed starts; since a merge is determined by these numbers, $\pi_t$ is realized, and no relative order of events of $G_{t-1}$ changes. The starts of $M_{t+1}$ are placed at the cursor like the points; their order relative to the ends of $M_{t+1}$ is fixed in the next step. The order $G_{R+1}$ is the required interleaving.
\end{proof}

\subsection{The dynamic program}\label{sec:dp}

The algorithm processes the events of all sequences one at a time, in an order it chooses, while maintaining the state
\[
(\nu_0,\dots,\nu_{R+1};\ s_1,\dots,s_{R+1};\ u_1,\dots,u_{R+1}),
\]
where $\nu_t$ is the number of processed events of $\sigma_t$, and $s_j$ and $u_j$ are the numbers of processed starts and processed ends of selected members of $M_j$. As for merges, a member $x$ is selected exactly when its start is processed before its end. This can be tested at either endpoint from $\nu$ alone, since the positions of the start and of the end of $x$ in their sequences are fixed: when the start of $x$ is processed, $x$ is selected if its end has not been processed yet, and when its end is processed, $x$ is selected if its start has been processed. A \emph{transition} processes the next event of some sequence $\sigma_t$:
\begin{itemize}
\item the start of a member $x\in M_{t+1}$: if $x$ is selected, require $s_{t+1}<a_{t+1}$ and increase $s_{t+1}$;
\item the end of a member $x\in M_t$: if $x$ is selected, require $u_t<a_t$ and increase $u_t$;
\item a point $p\in E_t$: gain $f_p\bigl(c_F(p)+(a_t-u_t)+s_{t+1}\bigr)$.
\end{itemize}
The initial state is zero everywhere; a state is \emph{accepting} if every sequence is exhausted and $s_j=u_j=a_j$ for all $j$. The following lemma explains the gain.

\begin{lemma}[two counters]\label{lem:counters}
Process the events along an interleaving that selects exactly $a_j$ members of each $M_j$, and let $W$ be the resulting window selection. When a point $p\in E_t$ is processed, $c_W(p)=c_F(p)+(a_t-u_t)+s_{t+1}$.
\end{lemma}

\begin{proof}
The events of $\sigma_t$ processed before $p$ are exactly those that precede $p$ in $\sigma_t$. By \Cref{lem:structure}(iii), the selected members of $M_t$ that contain $p$ are those whose ends follow $p$ in $\sigma_t$; their number is $a_t$ minus the number $u_t$ of selected ends preceding $p$. The selected members of $M_{t+1}$ that contain $p$ are those whose starts precede $p$ in $\sigma_t$, and $s_{t+1}$ counts exactly these.
\end{proof}

The prescribed size $a_t$ is essential here: the start of a selected member of $M_t$ lies in $\sigma_{t-1}$ and may be processed after $p$, so these members cannot be counted through their starts. For the same reason, the gains collected by a sequence of transitions are the true values only if it ends in an accepting state, and only such sequences are used.

\begin{proof}[Proof of \Cref{thm:window}]
Every sequence of transitions from the initial state to an accepting state processes the events along an interleaving. The selected sets are closed by \Cref{lem:merge}, each has size $a_j$ because every selected member increases $s_j$ exactly once, and the total gain is $\val(W)$ by \Cref{lem:counters}. Conversely, by \Cref{lem:closed-suffice} some optimal choice consists of closed sets; by \Cref{lem:merge} they are selected by merges, and by \Cref{lem:compatible} some interleaving realizes all these merges. Following this interleaving never violates a requirement and ends in an accepting state with total gain $\val(W)$.

Gains and feasibility of transitions depend only on the state, and every transition increases $\sum_t\nu_t$; so the best gain of every state is computed in topological order, and predecessor pointers recover an optimal choice. There are at most $(2n+m+1)^{R+2}(n+1)^{2R+2}$ states with at most $R+2$ transitions each, and every stored value is a sum of at most $m$ table entries.
\end{proof}
\section{From windows to an approximation scheme}\label{sec:ptas}

We combine windows and prove \Cref{thm:ptas}. The difficulty is the budget: windows taken from an optimal solution share long intervals that contain several representatives, so adding up their sizes would count such intervals repeatedly.

\paragraph*{The algorithm.}
Fix an integer $L\ge2$. For every $R\in\{1,\dots,L-1\}$, every sequence $A_1\prec\dots\prec A_{R+2}$ in $\cA$, and all integers $a_1,\dots,a_{R+1}$ with $0\le a_j\le\min\{|M_j|,K\}$, compute with \Cref{thm:window} the maximum value $\val^*$ of a choice and an optimal choice, together with the cost $\kappa$ of the window (\Cref{lem:structure}(i)). Build a directed graph whose nodes are the pairs $(A,k)$ with $A\in\cA$ and $0\le k\le K$. Every window contributes the arcs
\[
(A_1,k)\longrightarrow(A_{R+2},k+\kappa)\quad\text{of weight }\val^*,\qquad\text{for all $k$ with }k+\kappa\le K.
\]
The graph is acyclic because $r(A_{R+2})>r(A_1)$. A path is \emph{admissible} if it starts at a node $(A,|\Cl(A)|)$; its \emph{output} is the union of $\Cl(A)$ with the window selections of its arcs. The algorithm computes an admissible path of maximum weight and returns its output.

\begin{lemma}[soundness]\label{lem:sound}
The output $S$ of an admissible path that ends at $(A,k)$ satisfies $|S|\le k\le K$ and $f(S)$ is at least the weight of the path.
\end{lemma}

\begin{proof}
We show by induction along the path that the union built so far has at most as many intervals as the current second coordinate. This holds at the first node. Before an arc of a window with boundaries $A_1\prec\dots\prec A_{R+2}$ is used, the union contains $\Cl(A_1)$: either the arc is the first one and $\Cl(A_1)$ is the initial set, or the union contains the fixed set of the previous window, which includes the containers of that window's last boundary, and this boundary is $A_1$. The arc therefore adds at most $|W\setminus\Cl(A_1)|=\kappa$ new intervals.

For the value, observe that the kept slabs of consecutive windows on the path are disjoint: the kept slabs of a window end at $r(A_{R+1})$, whereas those of the next window begin at the right endpoint of its first boundary, which is $r(A_{R+2})$. Every point is therefore counted by at most one window. If a window counts $p$, its window selection $W$ is contained in $S$, so $c_W(p)\le c_S(p)$ and $f_p(c_W(p))\le f_p(c_S(p))$. Points counted by no window contribute nonnegative values to $f(S)$.
\end{proof}

\paragraph*{Windows from an optimal solution.}
Let $O$ and its representatives $I_0,\dots,I_{h+2}$ be as in \Cref{sec:prelim}. For consecutive slabs $\Sigma_\alpha,\dots,\Sigma_\beta$ with $1\le\alpha\le\beta\le h+1$ and $R=\beta-\alpha+1$, the \emph{optimal window} has boundaries $A_j=I_{\alpha-2+j}$ for $1\le j\le R+2$ and sizes $a_j=|O\cap M_j|$; its kept slabs are $\Sigma_\alpha,\dots,\Sigma_\beta$. Let $W^O=F\cup\bigcup_j(O\cap M_j)$.

\begin{lemma}\label{lem:optwindow}
The set $W^O$ is contained in $O$, each $O\cap M_j$ is closed, and $c_{W^O}(p)=c_O(p)$ for every point $p$ in $\Sigma_\alpha\cup\dots\cup\Sigma_\beta$. Hence the maximum value of a choice in the optimal window is at least $\sum_{p}f_p(c_O(p))$, summed over the points of these slabs.
\end{lemma}

\begin{proof}
The real boundaries are representatives and belong to $O$, so their containers belong to $O$ because $O$ is upward closed; sentinels have no containers. The same property makes $O\cap M_j$ closed. For a point $p\in\Sigma_i$ with $\alpha\le i\le\beta$, \Cref{lem:locality} shows that every interval of $O$ containing $p$ is a container of $I_{i-1}$, $I_i$ or $I_{i+1}$, all of which are boundaries of the window, or a member of $\Cr(I_{i-1},I_i)$ or $\Cr(I_i,I_{i+1})$, both of which are families of the window. Hence $W^O$ contains all intervals of $O$ that contain $p$.
\end{proof}

\paragraph*{Disjoint charges.}
Fix $\vartheta\in\{0,\dots,L-1\}$ and call a slab $\Sigma_i$ \emph{kept} if $i\not\equiv\vartheta\pmod L$. The kept slabs form maximal runs $[\alpha_1,\beta_1],\dots,[\alpha_\eta,\beta_\eta]$ of at most $L-1$ consecutive indices, with $\alpha_{\tau+1}=\beta_\tau+2$. Let $W^O_\tau$ be the set $W^O$ of the optimal window of the $\tau$-th run, and define the \emph{charged sets}
\[
D_0=\Cl(I_{\alpha_1-1}),\qquad D_\tau=W^O_\tau\setminus\Cl(I_{\alpha_\tau-1})\quad(1\le\tau\le\eta).
\]
Thus $|D_\tau|$ is the cost of the $\tau$-th optimal window.

\begin{lemma}[disjoint charges]\label{lem:charges}
The sets $D_0,D_1,\dots,D_\eta$ are pairwise disjoint subsets of $O$.
\end{lemma}

\begin{proof}
They are subsets of $O$ by \Cref{lem:optwindow}. Let $\tau<\tau'$ and $J\in W^O_\tau\cap W^O_{\tau'}$, and put $b=\alpha_{\tau'}-1$; then $b\ge\beta_\tau+1$. We show that $J$ contains $I_b$, so that $J\notin D_{\tau'}$. In the earlier window, $J$ is a container of some $I_c$ or a member of some $\Cr(I_{c-1},I_c)$, where $c\le\beta_\tau+1\le b$. In the later window, $J$ is a container of some $I_{c'}$ with $c'\ge b$ or a member of some $\Cr(I_{c'-1},I_{c'})$ with $c'-1\ge b$. A member of an earlier family ends before $r(I_c)\le r(I_b)$, whereas a container of $I_{c'}$ ends at or after $r(I_{c'})\ge r(I_b)$ and a member of a later family ends after $r(I_{c'-1})\ge r(I_b)$; hence $J$ is a container of $I_c$ in the earlier window. A container of $I_c$ starts at or before $\ell(I_c)\le\ell(I_b)$, whereas a member of a later family starts after $\ell(I_{c'-1})\ge\ell(I_b)$; hence $J$ is a container of $I_{c'}$ in the later window. Since $J$ contains $I_c$ and $I_{c'}$ with $c\le b\le c'$, and sentinels have no containers, $I_b$ is a representative and $J$ contains it (possibly $J=I_b$). The set $D_0$ consists of containers of $I_{\alpha_1-1}$, so the same argument, with $c=\alpha_1-1$, shows that $D_0$ is disjoint from every $D_{\tau'}$.
\end{proof}

\begin{theorem}\label{thm:shifting}
For every integer $L\ge2$, the algorithm returns a set $S$ with $|S|\le K$ and $f(S)\ge(1-1/L)\OPT$, using $(n+m+1)^{O(L)}$ operations on numbers of polynomial bit length.
\end{theorem}

\begin{proof}
The values $f_p(c_O(p))$ are nonnegative and sum to $\OPT$, and every point lies in exactly one slab; so for some $\vartheta$ the slabs with index congruent to $\vartheta$ modulo $L$ contribute at most $\OPT/L$. If no slab is kept, then $\OPT\le\OPT/L$, so $\OPT=0$ and any output suffices. Otherwise consider the admissible path that starts at $(I_{\alpha_1-1},|\Cl(I_{\alpha_1-1})|)$ and uses the arcs of the optimal windows of the runs, in order. Consecutive arcs fit together, because the last boundary $I_{\beta_\tau+1}$ of one run is the first boundary $I_{\alpha_{\tau+1}-1}$ of the next. The second coordinate of every node on the path is a partial sum of $|D_0|+|D_1|+\dots+|D_\eta|\le|O|\le K$ by \Cref{lem:charges}, so all these arcs exist. By \Cref{lem:optwindow}, the weight of the path is at least the contribution of the kept slabs, which is at least $(1-1/L)\OPT$. By \Cref{lem:sound}, the output of the maximum-weight path has $|S|\le K$ and value at least this weight.

The node $(\bot,0)$ starts an admissible path, so the algorithm always returns a set. There are at most $L(n+3)^{L+1}(n+1)^{L}$ windows, each solved with $O(L)\cdot(2n+m+1)^{L+1}(n+1)^{2L}$ operations by the proof of \Cref{thm:window}, which gives $O(L^2(2n+m+3)^{5L+2})$ operations in total and dominates the preprocessing and the path computation; the graph has $(n+3)(K+1)$ nodes and at most $K+1$ arcs per window, and path weights are sums of at most $m$ table entries.
\end{proof}

\begin{proof}[Proof of \Cref{thm:ptas}]
Apply \Cref{thm:shifting} with $L=\max\{2,\lceil1/\eps\rceil\}$.
\end{proof}
\section{General local rewards and dense subgraphs}\label{sec:general}

We extend the algorithm to the local rewards described in \Cref{sec:results} and derive \Cref{thm:dense}; for convex bipartite graphs, optimal solutions need not select consecutive vertices (\Cref{rem:nonconsecutive}).

\paragraph*{The reward model.}
Besides the intervals, the input may contain a finite set $V$ of \emph{optional points}, which can be selected as well. A solution is a pair $(S,S_V)$ with $S\subseteq\I$ and $S_V\subseteq V$, subject either to $|S|\le K$ and $|S_V|\le K_V$, or to $|S|+|S_V|\le K$. The objective is a sum of nonnegative rational rewards of three kinds, each given by an explicit table of its values for coverage $0,\dots,n$:
\begin{enumerate}[(i)]
\item for every point $p\in P$, a \emph{point reward} $f_p(c_S(p))$, where $f_p$ is nondecreasing;
\item for every interval $J\in\I$, an \emph{interval reward} $g_J(\chi_J,c_S(\ell(J)))$, where $\chi_J=1$ if $J\in S$ and $\chi_J=0$ otherwise, and $g_J$ is nondecreasing in both arguments;
\item for every optional point $v\in V$, an \emph{optional-point reward} $\varphi_v(\chi_v,c_S(v))$, where $\chi_v=1$ if $v\in S_V$ and $\chi_v=0$ otherwise, and $\varphi_v(\chi,\cdot)$ is nondecreasing for each $\chi\in\{0,1\}$; no relation between $\varphi_v(0,\cdot)$ and $\varphi_v(1,\cdot)$ is required.
\end{enumerate}
Here $c_S(\ell(J))$ counts the selected intervals that contain the start of $J$, excluding $J$.

\begin{theorem}[general approximation scheme]\label{thm:general}
Suppose that some optimal solution $(O,O_V)$ of an objective of this form has an upward-closed interval set $O$. Then for every integer $L\ge2$, a feasible solution of value at least $(1-1/L)\OPT$ can be computed with $(n+m+|V|+1)^{O(L)}$ arithmetic operations on numbers of polynomial bit length.
\end{theorem}

\begin{proof}[Proof sketch]
The algorithm and its analysis are those of \Cref{sec:window,sec:ptas} with three additions, detailed in \Cref{app:general}. First, the event sequence of a kept slab $E_t$ also contains its optional points and the starts of all intervals in $E_t$. Such an interval belongs to $F$, to $M_{t+1}$, or to no family, so its selection status is $1$, the verdict of the dynamic program, or $0$, and $c_W(\ell(J))$ is given by the formula of \Cref{lem:counters}, in which $s_{t+1}$ does not yet include $J$. Second, the program branches on the status of every optional point and counts the selected ones; this count becomes an additional coordinate of the graph of \Cref{sec:ptas}. Third, when windows are combined, coverage values and the status of intervals can only increase, while every optional point lies in the kept slabs of at most one window and keeps its status; this is where monotonicity is used, in particular that of $g_J$ in its first argument, because an interval counted with status $0$ by one window may be selected by another. \Cref{lem:closed-suffice} fails for interval rewards, since replacing a selected interval by a container can lower the reward of the replaced interval; the program still optimizes over closed choices only, which suffices because the choice of an optimal window is closed. In the analysis, \Cref{lem:locality} applied at the start of an interval shows that an optimal window contains every interval of $O$ that starts in its kept slabs.
\end{proof}

The applications below verify the hypothesis with \Cref{lem:exchange}.

\paragraph*{Convex bipartite graphs.}
Let $G=(U\cup V,E)$ be convex with respect to the order $v_1,\dots,v_{|V|}$ of $V$; if no such order is given, one can be computed in linear time by testing the consecutive-ones property~\cite{booth-lueker1976}. Place $v_i$ at position $i$ and represent every $u\in U$ with neighbourhood $\{v_a,\dots,v_b\}$ by the interval $[a-\frac12,b+\frac12]$; vertices of $U$ without neighbours affect no objective and are ignored. Then $c_S(v)=|N(v)\cap S|$ for every set $S$ of such intervals and every $v\in V$. With the optional points $V$ and the rewards $\varphi_v(\chi,d)=\chi\cdot d$, the objective of \Cref{thm:general} equals $|E(S,S_V)|$. Replacing a selected interval by an unselected container keeps $S_V$ and both budgets, and it does not decrease the coverage of any point, so \Cref{lem:exchange} provides an upward-closed optimal solution, and \Cref{thm:general} proves \Cref{thm:dense}(a).

\paragraph*{Interval graphs.}
Now the vertices are the intervals of $\I$, adjacent when they intersect (a representation can be computed in linear time from the graph~\cite{booth-lueker1976}), and we select $S\subseteq\I$ with $|S|\le K$. We charge each clique, and each edge, to its member that starts last.

\begin{lemma}[charging to the latest start]\label{lem:pivot}
For every $S\subseteq\I$ and every $r\ge2$,
\[
N_r(S)=\sum_{J\in S}\binom{c_S(\ell(J))}{r-1},
\]
and for rationals $0=\mu_0\le\mu_1\le\mu_2$ the edge utility of $S$ equals
\[
\sum_{J\in\I}\Bigl(\chi_J\bigl(\mu_1c_{\I}(\ell(J))+(\mu_2-\mu_1)\,c_S(\ell(J))\bigr)+(1-\chi_J)\,\mu_1\,c_S(\ell(J))\Bigr).
\]
\end{lemma}
\begin{proof}
Every member of a clique other than the one that starts last, say $J$, starts before $\ell(J)$ and intersects $J$, so it contains $\ell(J)$; conversely, any intervals that contain $\ell(J)$ form a clique together with $J$. This gives the first identity. For the second, charge every edge to its later-starting endpoint $J$; the earlier-starting neighbours of $J$ are exactly the $c_{\I}(\ell(J))$ intervals containing $\ell(J)$, of which $c_S(\ell(J))$ are selected. If $J$ is selected, these edges are worth $\mu_2$ each for selected and $\mu_1$ for unselected neighbours; if $J$ is not selected, they are worth $\mu_1$ for selected and $0$ for unselected neighbours.
\end{proof}
\begin{proof}[Proof of \Cref{thm:dense}(b) and~(c)]
For~(b), take $g_J(\chi,d)=\chi\sum_{r=2}^{K}\lambda_r\binom{d}{r-1}$ as interval rewards. For~(c), with $\bar d=\min\{d,c_{\I}(\ell(J))\}$, use $g_J(0,d)=\mu_1\bar d$ and $g_J(1,d)=\mu_1c_{\I}(\ell(J))+(\mu_2-\mu_1)\bar d$; the clipping changes no value that occurs. These rewards are nonnegative and nondecreasing in $d$, and $g_J(1,d)-g_J(0,d)$ equals $\sum_{r}\lambda_r\binom{d}{r-1}\ge0$ in~(b) and $\mu_1(c_{\I}(\ell(J))-\bar d)+(\mu_2-\mu_1)\bar d\ge0$ in~(c). By \Cref{lem:pivot} they sum to the objectives.

Let $I\in S$ and let $J'\notin S$ contain $I$. Every interval intersecting $I$ intersects $J'$, so replacing $I$ by $J'$ maps the cliques containing $I$ injectively to cliques containing $J'$. For the edge utility, the edge $\{I,J'\}$ keeps one selected endpoint, the edges from a common neighbour to $I$ and to $J'$ exchange their values, and an edge from $J'$ to a vertex not adjacent to $I$ gains $\mu_{\chi+1}-\mu_\chi\ge0$, where $\chi$ is the status of that vertex. The budget is unchanged, so neither objective decreases.
Replacing a selected interval by an unselected container does not decrease either objective, so \Cref{lem:exchange,thm:general} apply.
\end{proof}
\section{NP-hardness}\label{sec:hardness}

We prove \Cref{thm:hardness} in two stages. First, we establish strong NP-hardness of Maximum Exact Interval Multicover (\MaxEIMC{}) on proper interval families by a reduction from Rectilinear Steiner Arborescence (RSA), using a bounded integral flow problem on a directed grid as an intermediate step. Second, we reduce \MaxEIMC{} to \MaxIMC{}, ensuring that all intervals have equal length.
In \MaxEIMC{}, we are given a family \(\mathcal I\) of intervals, a set $P$ of points, each with a target \(z_p\in\mathbb Z_{\ge 0}\) and a positive weight \(\lambda_p\), and an integer \(K\).
Each point $p$ contributes \(\lambda_p\) to the objective if \(c_S(p)=z_p\), and \(0\) otherwise. The objective is to select exactly \(K\) intervals maximizing the total weight of satisfied points. A family of intervals is proper if its endpoints are distinct and its starts appear in the same order as its ends; for distinct endpoints, this means that no interval contains another.
%A family of intervals is \emph{proper} if its endpoints are distinct and its starts appear in the same order as its ends; for distinct endpoints, this means that no interval contains another.

\subsection{From Steiner arborescences to exact-coverage tests}\label{sec:rsa}

In the \emph{rectilinear Steiner arborescence} problem (RSA), we are given a set $Z$ of terminals in the plane, each with with nonnegative integer coordinates. We seek a shortest union of horizontal and vertical segments containing, for every terminal, a path from the origin $o$ that is monotone in both coordinates.
RSA is strongly NP-complete~\cite{shi-su2006} (see also~\cite[Section~2]{aichholzer-mulzer-pilz2015}), so we may assume that all coordinates are bounded by a polynomial in $|Z|$; we also assume $o\notin Z$. Let \(\Gamma\) denote the directed integer grid on the smallest rectangle containing \(o\) and \(Z\), with arcs pointing rightward and upward.
By a standard observation [21], \(\mathrm{OPT}_{\mathrm{RSA}}\) equals the minimum number of arcs in an arborescence of \(\Gamma\) rooted at \(o\) that contains \(Z\).
%Let $e$ be the number of arcs of $\Gamma$ and $B=|Z|$. 
By \Cref{lem:flow}, RSA reduces to the following bounded integral flow problem on the directed grid \(\Gamma\): minimize the number of arcs with positive flow, subject to \(0\le y_a\le B\) for every arc \(a\), supply \(B=|Z|\) at \(o\), demand one at each terminal, and zero supply elsewhere. Such flows exist, and $B\le|V(\Gamma)|$.
Equivalently, we may maximize the number of arcs carrying zero flow.
We denote the optimum of this maximization problem by $\mathrm{OPT}_\text{flow} = \lvert E(\Gamma)\rvert - \mathrm{OPT}_\text{RSA}$. 

\paragraph*{The construction.}
We now reduce this bounded integral flow problem on a directed grid to \MaxEIMC{}. The grid order below is topological and admits consecutive frontier replacements; these are the structural properties used by the construction.
Process the vertices of $\Gamma$ row by row from the bottom and, within a row, from left to right. 
The \emph{frontier} consists of the arcs from processed to unprocessed vertices.
By \Cref{lem:frontier}, these arcs can be ordered such that processing a vertex \(v\) replaces a consecutive subsequence \(\operatorname{In}(v)\) of incoming arcs by the outgoing arcs \(\operatorname{Out}(v)\). More precisely, the ordered frontier changes from \(\Phi'\operatorname{In}(v)\Phi''\) to \(\Phi'\operatorname{Out}(v)\Phi''\), where juxtaposition denotes list concatenation. After every vertex we insert an \emph{identity stage}, which leaves the frontier unchanged. This gives stages $t=0,1,\dots,2|V(\Gamma)|$ with frontiers $\Phi_t$, the first and the last of them empty. Let $z_t$ be the total supply of the vertices processed by stage $t$, with supply $B$ at $o$, $-1$ at terminals and $0$ elsewhere; for every feasible flow, the total flow on $\Phi_t$ is $z_t$.

For every stage $t$, arc $a\in\Phi_t$ and $j\in\{1,\dots,B\}$ we create a \emph{packet} of \(B\) identical candidate intervals. 
The number of selected candidates from this packet represents the flow on \(a\). Every stage $t$ has a \emph{center} site. The packets of stage $t$ start between the centers of stages $t-1$ and $t$ and end between the centers of stages $t$ and $t+1$, so the center of stage $t$ is covered by exactly the candidates of stage $t$. 
%Ending or starting a packet places the endpoints of its candidates consecutively, in the order of $j$.
Between two consecutive centers, the packets of stage $t-1$ end and those of stage $t$ start in the following order, interleaved with points of two kinds: \emph{guards}, of weight $\Lambda=|E(\Gamma)|+1$, and \emph{probes}, of weight $1$.
Intuitively, guards enforce the flow's feasibility, whereas probes reward arcs carrying no flow.
Our choice of weights will guarantee that any optimal solution must satisfy all guards.
\begin{itemize}
\item The center of stage $t$ carries a guard with target $z_t$.
\item If stage $t$ processes $v$: for each arc of $\Phi'$ in order, end its old packet, start its new packet, and place a guard with target $z_{t-1}$; then end the packets of $\mathrm{In}(v)$, start those of $\mathrm{Out}(v)$, and place a guard with target $z_t$; then treat $\Phi''$ like $\Phi'$, with target $z_t$.
\item If stage $t$ is an identity stage: for each frontier arc in order, end its old packet, start its new packet, and place a guard with target $z_t$; if the arc was created in the preceding vertex stage, place a probe with target $z_t$ between its two packets.
\end{itemize}
Finally, we slightly perturb the endpoints of the candidate intervals in each packet to obtain a proper interval family. 
Sites and endpoints receive consecutive integer coordinates in this order. Let $K=\sum_tz_t$, let $g$ be the number of guards and $\Psi(S)$ the total weight of the tests satisfied by $S$.

\begin{figure}[t]
\centering
\scalebox{1.074809}{%
\begin{tikzpicture}[x=1cm,y=0.26cm,font=\footnotesize]
  \begin{scope}[scale=1.328,xscale=1.075,transform shape]
  \foreach \x/\lab in {0.28/{center $t{-}1$},2.65/{center $t$},6.5305/{center $t{+}1$}}{
    \draw[black,thin] (\x,-0.9)--(\x,6.5); \node[above] at (\x,6.4) {\lab};}
  \node[above] at (1.43324,8.0) {vertex stage $t$};
  \node[above] at (4.6026,8.0) {identity stage $t{+}1$};
  \foreach \x in {0.98,1.65,2.25,3.2,4.3973,5.5092,6.1156}{\draw[black,dashed] (\x,-0.9)--(\x,6.0);}
  \foreach \x in {3.9133,4.9227}{\draw[famorange,densely dotted,thick] (\x,-0.9)--(\x,6.0);}
  \draw[->,gray] (-0.04,-0.9)--(6.9515,-0.9);
  \def\pk#1#2#3#4{\draw[line width=2.3pt,#4] (#1,#3)--(#2,#3);}
  \pk{0}{0.59818}{5}{gray!55}\pk{0.78909}{2.89627}{5}{selblue}
  \pk{3.09328}{6.9128}{5}{famteal}
  \pk{0}{1.055}{4}{gray!55}
  \pk{0}{1.17451}{3}{gray!55}
  \pk{1.29402}{3.395}{2}{selblue}\pk{4.2763}{6.9128}{2}{famteal}
  \pk{1.41353}{4.49284}{1}{selblue}\pk{5.32874}{6.9128}{1}{famteal}
  \pk{0}{1.87105}{0}{gray!55}\pk{2.06053}{5.70012}{0}{selblue}\pk{5.90786}{6.9128}{0}{famteal}
  \foreach \y/\lab in {5/a,4/b,3/{b'},2/c,1/{c'},0/d}{\node[left] at (0,\y) {$\lab$};}
  \foreach \x/\lab in {0.28/{z_{t-1}},0.98/{z_{t-1}},1.65/{z_t},2.25/{z_t},2.65/{z_t}}
  {\node[below] at (\x,-0.9) {$\lab$};}
  \foreach \x in {3.2,4.3973,5.5092,6.1156}
    {\node[below,font=\scriptsize] at (\x,-0.9) {$z_{t+1}$};}
  \foreach \x in {3.9133,4.9227}
    {\node[below,font=\scriptsize] at (\x,-1.8) {$z_{t+1}$};}
  \end{scope}
  \begin{scope}[shift={(11.1583,2.8)},x=1.35cm,y=1.35cm,font=\fontsize{11}{13}\selectfont]
    \draw[-{Stealth[length=3mm,width=1.6mm]},gray!80,line width=1.2pt] (-0.8,0)--(-0.07,0) node[midway,above] {$b$};
    \draw[-{Stealth[length=3mm,width=1.6mm]},gray!80,line width=1.2pt] (0,-0.7)--(0,-0.07) node[midway,left] {$b'$};
    \draw[-{Stealth[length=3mm,width=1.6mm]},selblue,line width=1.2pt] (0,0.07)--(0,0.7) node[midway,left] {$c$};
    \draw[-{Stealth[length=3mm,width=1.6mm]},selblue,line width=1.2pt] (0.07,0)--(0.8,0) node[midway,above] {$c'$};
    \fill (0,0) circle (3.2pt);
    \node[below right,inner sep=3pt,font=\fontsize{13}{15}\selectfont] at (0,0) {$v$};
  \end{scope}
\end{tikzpicture}%
}
\caption{Stages $t$, which processes $v$ with $\mathrm{In}(v)=(b,b')$ and $\mathrm{Out}(v)=(c,c')$, and $t+1$, an identity stage, between the centers of stages $t-1$ and $t+1$, in the order of endpoints before lengths are adjusted. The ordered frontier changes from $\Phi_{t-1}=(a,b,b',d)$ to $\Phi_t=(a,c,c',d)$: the consecutive block $(b,b')$ is replaced by $(c,c')$, preserving the order of the remaining arcs. The identity stage leaves $\Phi_{t+1}=\Phi_t$. Packet starts and ends follow these orders, ensuring properness. The sketch on the right shows the arcs incident to $v$. Bars are packets of stage $t-1$ (gray), $t$ (blue) and $t+1$ (green); dashed lines are guards, dotted lines probes, and their respective targets are shown at the bottom.
If all guards are satisfied, every arc keeps its number of selected candidates.
Furthermore, the two leftmost guards together enforce flow conservation at $v$, since $z_t-z_{t-1}$ is precisely the supply of $v$. The leftmost probe is satisfied exactly when $c$ carries no flow.}
\label{fig:packet}
\end{figure}
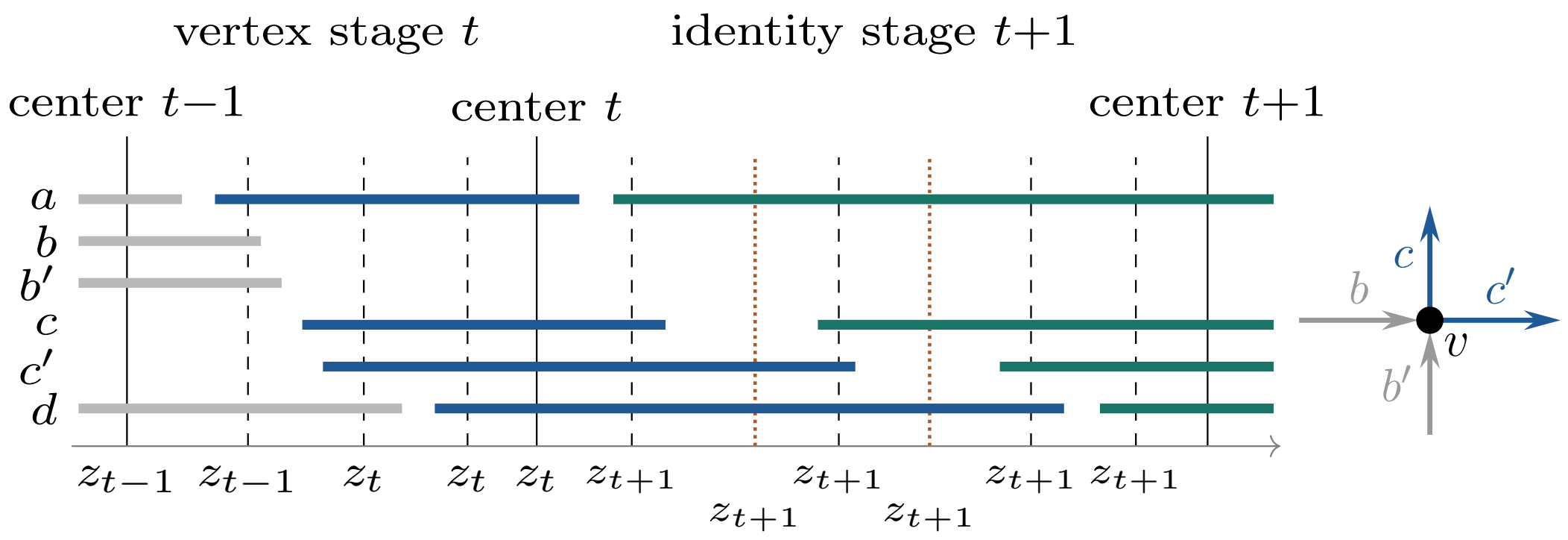

\begin{lemma}[exact-coverage instance]\label{lem:grid}
The construction takes polynomial time and yields a proper interval family, polynomially bounded positive integer point weights, and $K$ at most the number of candidates, such that
\(
\max_{|S|=K}\Psi(S)=\Lambda g +\OPT_{\mathrm{flow}}.
\)
\end{lemma}

\begin{proof}[Proof sketch]
Between the centers of stages $t-1$ and $t$, only candidates of these two stages are active, and at the center of stage $t$ exactly those of stage $t$. So if $S$ satisfies all guards, the center guards force $z_t$ selected candidates in each stage $t$, and $|S|=K$. A packet replacement changes the coverage by the difference of the old and new selected counts, and consecutive guards with equal targets force equal counts, so flow values are copied (\Cref{fig:packet}); replacing $\mathrm{In}(v)$ by $\mathrm{Out}(v)$ changes the coverage by $z_t-z_{t-1}$, the supply of $v$, so flow is conserved. Hence the counts form a feasible flow, and conversely every feasible flow is realized while satisfying all guards. A probe sees $z_t$ minus the count of its arc's old packet, so it rewards exactly the arcs without flow. 
Selections satisfying all guards thus reach $\Lambda g+\OPT_{\mathrm{flow}}$, while violating a guard leaves at most $\Lambda(g-1)+|E(\Gamma)|<\Lambda g$. In every stage, starts and ends of candidates both follow the frontier order and then $j$, so the family is proper. \Cref{app:hardness} gives the details.
\end{proof}

\subsection{From exact coverage to demands}\label{sec:tilt}

The existence of positive equal masses follows from the classical proper-to-unit interval conversion, which preserves endpoint order~\cite{bogart-west1999}.
Our reduction requires the following polynomially-bounded version.

\begin{lemma}[equal masses]\label{lem:masses}
Consider a family of $n_0$ intervals with distinct endpoints and $m_0$ points, none of them an endpoint, such that every gap between two consecutive endpoints contains a point. If the family is proper, there are integers $\omega_x\ge1$ for the points $x$ and $C\ge1$, computable in polynomial time, such that every interval $I$ of the family satisfies $\sum_{x\in I}\omega_x=C$, and $\omega_x\le n_0m_0+1$ and $C\le n_0m_0$. Conversely, if positive masses with equal sums exist, the family is proper.
\end{lemma}

We prove \Cref{lem:masses}. For a proper family, we first construct positive real masses using a strictly increasing cumulative-weight function whose increase over each interval is $1$. Total unimodularity then yields integer masses with polynomial bounds.

Assume first that the family is proper. Index the intervals $J_1,\dots,J_{n_0}$ by their starts and the points $x_1<\dots<x_{m_0}$ from left to right. Every interval contains a point; let $\alpha_i$ and $\beta_i$ denote the indices of the first and last points in $J_i$. We have $\alpha_1<\dots<\alpha_{n_0}$ and $\beta_1<\dots<\beta_{n_0}$ by properness and the presence of a point in every gap between consecutive endpoints.

\paragraph*{Real masses.} We construct a strictly increasing $\Pi:\{0,\dots,m_0\}\to\mathbb{R}$ with $\Pi(\beta_i)-\Pi(\alpha_i-1)=1$ for all $i$. Here $\Pi(k)$ represents the cumulative weight of the first $k$ points. Let $\beta_0=0$ and $\Pi(0)=0$, and handle $i=1,2,\dots$ in turn, with $\Pi$ defined on $\{0,\dots,\beta_{i-1}\}$: if $\alpha_i-1>\beta_{i-1}$, extend $\Pi$ strictly increasingly up to $\alpha_i-1$; then set $\Pi(\beta_i)=\Pi(\alpha_i-1)+1$ and interpolate linearly between $\max\{\beta_{i-1},\alpha_i-1\}$ and $\beta_i$. This keeps $\Pi$ increasing, because $\Pi(\beta_i)>\Pi(\alpha_i-1)$ and, if $\alpha_i-1<\beta_{i-1}$, also $\Pi(\beta_i)>\Pi(\alpha_{i-1}-1)+1=\Pi(\beta_{i-1})$ since $\alpha_i>\alpha_{i-1}$. Finally, extend $\Pi$ strictly increasingly up to $m_0$. The differences $\Pi(k)-\Pi(k-1)$ are positive masses with all interval sums $1$; after scaling, $\omega\ge1$.

\paragraph*{Integral masses.} Let $\mathbf{A}$ be the $n_0\times m_0$ incidence matrix with $A_{ih}=1$ if and only if $\alpha_i\le h\le\beta_i$. If $A_{ih}=A_{kh}=1$ for $i<j<k$, then $\alpha_j<\alpha_k\le h\le\beta_i<\beta_j$, so $A_{jh}=1$ as well. Thus every column has consecutive ones. Appending a column of ones preserves this property, and a matrix with consecutive ones in every column is totally unimodular, as is well known. (In a square submatrix, subtracting from each row its predecessor, from the bottom up, preserves the determinant and leaves in each column at most one entry $+1$ and at most one entry $-1$, which gives the incidence matrix of a directed graph.) Negating the appended column preserves total unimodularity, so the matrix $[\mathbf{A}\mid-\one]$ is totally unimodular.

The real masses constructed above show that the polyhedron $\{(\zeta,C)\ge0:\ [\mathbf{A}\mid-\one](\zeta,C)^{\top}=-\mathbf{A}\one\}$ is nonempty, where $\omega=\one+\zeta$. After deleting redundant equations, it has a basic feasible solution, which can be computed in polynomial time by linear programming. Its basis matrix has at most $n_0$ rows, determinant $\pm1$, and cofactors in $\{0,\pm1\}$. By Cramer's rule, every basic variable equals a determinant obtained from the basis matrix by replacing one column with the right-hand side, whose entries have absolute value at most $m_0$; expanding along this column bounds the variable by $n_0m_0$. Hence $\zeta$ and $C$ are integers bounded by $n_0m_0$, so $1\le\omega_x\le n_0m_0+1$, and $C\ge1$ because every interval contains a point.

\paragraph*{Converse.} If an interval strictly contains another, the gap immediately after its left endpoint contains a point outside the smaller interval. Positive point masses therefore give the larger interval strictly greater total mass, contradicting equal sums. \qed

We now reduce \MaxEIMC{} on proper interval families with polynomially bounded positive integer weights to \MaxIMC{}. Consider such an instance with $0\le K\le|\I|$, and let $\Lambda\ge1$ be an integer upper bound on its point weights. Add a point with zero reward in every gap between consecutive endpoints, and let $\omega$ and $C$ be as in \Cref{lem:masses}. For each original point $x$, set $\psi_x(d)=\lambda_x\one[d=z_x]$; for added points, set $\psi_x\equiv0$. Thus $\Psi(S)=\sum_x\psi_x(c_S(x))$. Define $T_x(d)=\Lambda\,\omega_x\,d+\psi_x(d)-\psi_x(0)$ for $0\le d\le K$. Then $T_x(0)=0$, and $T_x$ is nondecreasing because $\psi_x$ decreases by at most $\Lambda$ in one step while $\Lambda\omega_x\ge\Lambda$. Since every interval has mass $C$, double counting gives, for every $S$ with $|S|\le K$,
\begin{equation}\label{eq:tilt}
\sum_xT_x\bigl(c_S(x)\bigr)=\Lambda\sum_x\omega_x\,c_S(x)+\Psi(S)-\Psi(\emptyset)=\Lambda C|S|+\Psi(S)-\Psi(\emptyset).
\end{equation}
Replace every point $x$ by $T_x(d)-T_x(d-1)$ points of unit profit and demand $d$, for $d=1,\dots,K$, placed near $x$; their total profit at coverage $c\le K$ is $T_x(c)$. Assigning consecutive segments of lengths proportional to $\omega_x$ to the original and added points makes all intervals equally long (\Cref{app:tilt}). The number of satisfied points is nondecreasing in $S$ and $K\le|\I|$, so its maximum over $|S|\le K$ is attained at $|S|=K$. By \eqref{eq:tilt}, we have $\OPT_{\text{\MaxIMC}}=\Lambda CK-\Psi(\emptyset)+\OPT_{\text{\MaxEIMC}}$.
\Cref{app:tilt} verifies the size of the construction, its integer coordinates
of polynomial size, and the remaining conditions of \Cref{thm:hardness}.
Composed with the first reduction, this proves \Cref{thm:hardness};
membership in NP is clear.

\section{An exact algorithm for bounded demands}\label{sec:bounded}

\begin{theorem}\label{thm:bounded}
If all demands are at most $q_{\max}\ge1$, then \MaxIMC{} with profits can be solved exactly with $O\bigl((n+m)(K+1)(q_{\max}+1)\sum_{i=0}^{q_{\max}}\binom ni\bigr)$ operations besides sorting and arithmetic on profits.
\end{theorem}

\begin{proof}
Sweep the starts, points and ends in increasing order, with ties refined as in \Cref{sec:prelim}. A state $(k,H)$ records the number $k$ of intervals selected so far and a set $H$ of at most $q_{\max}$ selected intervals that are currently active; the table stores the largest profit of a history leading to the state. At a start, either skip the interval or, if $k<K$, select it: increase $k$ and insert the interval into $H$, and if $H$ then has $q_{\max}+1$ members, discard the one that ends first. At an end, remove the interval from $H$ if it is there. At a point $p$, gain $w_p$ if $|H|\ge q_p$.

We claim that $|H|=\min\{q_{\max},c_S(p)\}$ at every point $p$, where $S$ is the set of intervals selected so far. When an interval $x$ is discarded, $H$ consists of $q_{\max}$ selected intervals that end after $x$. As long as $x$ is active, none of them ends, and a later discard keeps $H$ a set of $q_{\max}$ intervals ending after $x$, because it removes the member of $H\cup\{y\}$ that ends first, where $y$ is the new interval. Hence $|H|=q_{\max}$ while a discarded interval is active, and when $|H|<q_{\max}$, every discarded interval has ended and $H$ consists of all active selected intervals. Since all demands are at most $q_{\max}$, the gains are exact. The future evolution depends only on $(k,H)$, so histories with the same state can be merged. There are at most $(K+1)\sum_{i\le q_{\max}}\binom ni$ states per event, and each update takes $O(q_{\max}+1)$ operations.
\end{proof}

The running time is exponential in $q_{\max}$, so this is an XP algorithm with respect to the maximum demand, not a fixed-parameter one.

\section{Concluding remarks}\label{sec:conclusion}
Open questions include an efficient PTAS for \MaxIMC{}, its complexity for a common demand given as part of the input, fixed-parameter tractability in the maximum demand, the complexity of Bipartite Dense Subgraph on convex bipartite graphs and of Maximum $k$-Vertex Cover on interval graphs, interval costs, under which containment exchange fails, and circular arcs, where crossing families form a cycle.

\paragraph*{Use of generative AI.}
The authors used OpenAI ChatGPT and Anthropic Claude during the research and writing process. The tools suggested formulations of 
definitions, theorem and proof drafts, leads for related-work searches, and edits to prose and organization; some 
suggestions were incorporated after human review and independent verification against the mathematics and primary literature. The 
authors assume responsibility for all content.

\clearpage
\bibliography{references}

\clearpage
\appendix
\section{Proof of the general approximation scheme}\label{app:general}

We prove \Cref{thm:general}. Recall the reward model of \Cref{sec:general}. All tables are extended to $\{0,\dots,n\}$ by saturation, which changes no value of a feasible solution.

\paragraph*{Windows with general rewards.}
Fix a window as in \Cref{def:window} and an integer $b\ge0$. A \emph{general choice} consists of sets $Y_j\subseteq M_j$ with $|Y_j|=a_j$, each of them closed, and a set $Y_V$ of exactly $b$ optional points in the kept slabs. With $W=F\cup Y_1\cup\dots\cup Y_{R+1}$, its value is the sum of the point rewards $f_p(c_W(p))$ and the optional-point rewards $\varphi_v(\one[v\in Y_V],c_W(v))$ over the points and optional points in the kept slabs, and of the interval rewards $g_J(\one[J\in W],c_W(\ell(J)))$ over the intervals $J$ that start in a kept slab.

\begin{lemma}\label{lem:general-window}
For every window and every $b$, a general choice of maximum value, and its value, can be computed with $(n+m+|V|+1)^{O(R)}$ arithmetic operations.
\end{lemma}

\begin{proof}
For $1\le t\le R$, the event sequence $\sigma_t$ now also contains the optional points of $E_t$ and the starts of all intervals of $\I$ that start in $E_t$; by \Cref{lem:structure}(ii), such an interval belongs to $F$, to $M_{t+1}$, or to no family, and only members of $M_{t+1}$ were already present. The state receives one more counter $b'\in\{0,\dots,b\}$, the number of selected optional points processed so far. The transitions of \Cref{sec:dp} are extended as follows; in all cases the gain is evaluated before the counters are updated.
\begin{itemize}
\item At an optional point $v\in E_t$, choose $\chi\in\{0,1\}$; if $\chi=1$, require $b'<b$ and increase $b'$. Gain $\varphi_v(\chi,c_F(v)+(a_t-u_t)+s_{t+1})$.
\item At the start of an interval $J$ in $E_t$, let $\chi=1$ if $J\in F$, let $\chi$ be the verdict of \Cref{sec:dp} if $J\in M_{t+1}$, and let $\chi=0$ otherwise. Gain $g_J(\chi,c_F(\ell(J))+(a_t-u_t)+s_{t+1})$, and then, if $J\in M_{t+1}$, update $s_{t+1}$ as before.
\end{itemize}
Accepting states additionally require $b'=b$. The proof of \Cref{lem:counters} applies verbatim to every position $x$ of $\sigma_t$, not only to points: the selected members of $M_t$ that cover $x$ are those whose ends follow $x$ in $\sigma_t$, and the selected members of $M_{t+1}$ that cover $x$ are those whose starts precede $x$ in $\sigma_t$. In particular, at the start of $J\in M_{t+1}$, the counter $s_{t+1}$ does not include $J$, in accordance with the definition of $c_W(\ell(J))$. The selection status used for $J$ is its status in $W$. Hence every accepting transition sequence yields a general choice whose value is the total gain; the selected sets are closed by \Cref{lem:merge}. Conversely, every general choice is realized by some interleaving (\Cref{lem:merge,lem:compatible}; the proof of \Cref{lem:compatible} applies verbatim, since the new events of $\sigma_t$ are placed at the cursor like points) together with the branch decisions $\chi_v=\one[v\in Y_V]$. If no general choice exists, no state is accepting and the window yields no arc. The state space grows by the factor $b+1\le|V|+1$.
\end{proof}

The dynamic program optimizes over closed choices only. This suffices, because the windows used in the analysis come from an upward-closed optimal solution, whose intersections with the families are closed.

\paragraph*{The algorithm.}
For every window and every $b$ with $0\le b\le|V|$, compute the maximum value $\val^*$ of a general choice by \Cref{lem:general-window}, together with the cost $\kappa$ of the window. The nodes of the graph are triples $(A,k,k_V)$ with $A\in\cA$, $0\le k\le K$ and $0\le k_V\le K_V$; under a total budget, the nodes satisfy $k+k_V\le K$ instead. A window with value $\val^*$ for $b$ yields the arcs $(A_1,k,k_V)\to(A_{R+2},k+\kappa,k_V+b)$ of weight $\val^*$ whenever the head is a node. Admissible paths start at nodes $(A,|\Cl(A)|,0)$, and the output of a path consists of the union of $\Cl(A)$ with the window selections of its arcs, together with the union of the sets $Y_V$ of its arcs.

\paragraph*{Soundness.}
Let $(S,S_V)$ be the output of an admissible path ending at $(A,k,k_V)$. As in \Cref{lem:sound}, $|S|\le k$. The kept slabs of the windows on the path are disjoint, so the sets $Y_V$ are disjoint and $|S_V|=k_V$; in particular the output is feasible. Every reward event, that is, every point, optional point and interval start, lies in the kept slabs of at most one window of the path. Consider an event counted by a window with selection $W\subseteq S$. A point reward does not decrease, because $c_W(p)\le c_S(p)$. An interval reward of $J$ does not decrease, because $\one[J\in W]\le\one[J\in S]$, $c_W(\ell(J))\le c_S(\ell(J))$, and $g_J$ is nondecreasing in both arguments. An optional point $v$ in the kept slabs of the window is selected in $S_V$ exactly when it belongs to $Y_V$, because no other window selects it; since $\varphi_v(\chi,\cdot)$ is nondecreasing, its reward does not decrease. Events counted by no window contribute nonnegative rewards. Hence the value of the output is at least the weight of the path.

\paragraph*{Approximation.}
Fix an optimal solution $(O,O_V)$ with $O$ upward closed, and define representatives and slabs as in \Cref{sec:prelim}. Every reward event lies in exactly one slab, because points, optional points and starts of intervals are distinct from the right endpoints of the representatives. Choose the residue $\vartheta$ as in the proof of \Cref{thm:shifting}, so that the rewards of $O$ in the slabs with index congruent to $\vartheta$ modulo $L$ sum to at most $\OPT/L$. For each run of kept slabs, take the optimal window of \Cref{sec:ptas} with $b=|O_V\cap(\Sigma_\alpha\cup\dots\cup\Sigma_\beta)|$ and the general choice $(O\cap M_j)_j$, $Y_V=O_V\cap(\Sigma_\alpha\cup\dots\cup\Sigma_\beta)$; it is closed because $O$ is upward closed. By \Cref{lem:locality}, applied at every position $x$ in the kept slabs, $W^O$ contains every interval $I\in O$ with $\ell(I)\le x\le r(I)$; hence $c_{W^O}(x)=c_O(x)$ for all points, optional points and starts in the kept slabs, and every interval of $O$ that starts in the kept slabs belongs to $W^O$. Intervals outside $O$ do not belong to $W^O\subseteq O$. Therefore this general choice has the same rewards as $(O,O_V)$ on the kept slabs. The interval coordinates along the resulting path are bounded by $|O|$ as in \Cref{lem:charges}, the optional-point coordinates by $|O_V|$ because the kept slabs of the runs are disjoint, and under a total budget their sum is at most $|O|+|O_V|\le K$; so all arcs exist. The path has weight at least $(1-1/L)\OPT$, and soundness transfers this weight to the output.

\paragraph*{Running time.}
We may assume $K\le n+|V|$. There are at most $L(n+3)^{L+1}(n+1)^{L}(|V|+1)$ pairs of windows and values of $b$, each solved with $(n+m+|V|+1)^{O(L)}$ operations, and the graph has polynomially many nodes and arcs. Every stored value is a sum of at most $m+|V|+n$ table entries. \qed

\begin{remark}[non-consecutive optima]\label{rem:nonconsecutive}
Optimal solutions need not select consecutive vertices of $U$: for neighbourhoods $\{v_1,v_2\}$, $\{v_3\}$ and $\{v_5,v_6\}$ and budgets $K_U=2$, $K_V=4$, the unique optimum selects the first and the third vertex. Since for fixed $S_U$ the best $S_V$ consists of the $K_V$ points with the most neighbours in $S_U$, and symmetrically, the difficulty lies entirely in the joint choice.
\end{remark}

\paragraph*{Why later endpoints.} Charging edges to their selected endpoints instead of their later-starting endpoints would give the term $\mu_1\deg(J)+(\mu_2-2\mu_1)\,c_S(\ell(J))$ for $J\in S$, which decreases in the coverage when $\mu_2<2\mu_1$ (as for Maximum $k$-Vertex Cover); such rewards are not monotone, and \Cref{lem:sound} would fail.

\section{Details of the hardness reduction}\label{app:hardness}

This appendix completes the proofs of \Cref{lem:grid} and \Cref{thm:hardness}. We use the notation of \Cref{sec:hardness}: $\Gamma$ is the directed grid on the smallest integer rectangle containing $o$ and $Z$, with $e$ arcs; $B=|Z|$; the supply is $B$ at $o$, $-1$ at the terminals and $0$ elsewhere.

\subsection{Flows and arborescences}

\begin{lemma}\label{lem:flow}
The minimum number of arcs of an arborescence in $\Gamma$ rooted at $o$ that contains $Z$ equals the minimum number of arcs with positive flow in an integral flow $y$ with $0\le y\le B$ whose net outflow at every vertex equals its supply. Such flows exist.
\end{lemma}

\begin{proof}
Given an arborescence, send one unit from $o$ to every terminal along its tree path; the flow is at most $B$ on every arc and positive only on arcs of the arborescence. Conversely, consider a feasible flow and a vertex $v\ne o$ incident to an arc with positive flow. If the arc leaves $v$, the inflow of $v$ is at least its outflow, because the supply of $v$ is nonpositive; so $v$ has an incoming arc with positive flow in either case, and so does every terminal. Choose one such incoming arc for every such vertex $v$. Following chosen arcs backwards never repeats a vertex, because $\Gamma$ is acyclic, and can only stop at $o$; hence the chosen arcs form an arborescence rooted at $o$ that contains $Z$, and it has at most as many arcs as the flow has arcs with positive flow. Feasible flows exist because monotone paths from $o$ to every terminal exist in the rectangle.
\end{proof}

\subsection{The frontier}

\begin{lemma}\label{lem:frontier}
When the vertices of $\Gamma$ are processed row by row from the bottom and, within each row, from left to right, the frontier can be listed so that processing a vertex $v$ replaces a consecutive block $\mathrm{In}(v)$, consisting of the arcs entering $v$ from the left and from below in this order, by the block $\mathrm{Out}(v)$, consisting of the existing arcs leaving $v$ upwards and to the right in this order.
\end{lemma}

\begin{proof}
Immediately before the vertex $v=(x,y)$ is processed, list the frontier as follows: first the upward arcs leaving the processed vertices of row $y$, from left to right; then the arc from $(x-1,y)$ to $v$, if it exists; finally the upward arcs leaving row $y-1$ at the columns $x,x+1,\dots$, from left to right. These are exactly the arcs from processed to unprocessed vertices. The arc entering $v$ from below is the first arc of the last group, so $\mathrm{In}(v)$ is consecutive. Replacing it by the upward arc leaving $v$ and the arc from $v$ to $(x+1,y)$ gives the same description for the next vertex of the row. At the end of a row, the frontier consists of the upward arcs leaving that row, which is the description for the first vertex of the next row. The first frontier is empty, so induction proves the claim; arcs that do not exist at the boundary of the rectangle are simply omitted.
\end{proof}

\subsection{Proof of \texorpdfstring{\Cref{lem:grid}}{the exact-coverage lemma}}

The stages, frontiers $\Phi_t$, totals $z_t$, packets, centers, guards and probes are as in \Cref{sec:rsa}. Summing the supplies of the processed vertices shows $z_t\in\{0,\dots,B\}$, $z_0=z_{\mathrm{last}}=0$, and, for every feasible flow, that the total flow on $\Phi_t$ is $z_t$, because no arc enters a processed vertex from an unprocessed one. For a selection $S$, let $x_{t,a}$ denote the number of selected candidates of the packet of $a$ in stage $t$.

\begin{claim}\label{cl:guards}
If $S$ satisfies all guards, then $|S|=K$, and there is a feasible flow $y$ with $x_{t,a}=y_a$ for all $t$ and $a\in\Phi_t$.
\end{claim}

\begin{proof}
Between the centers of stages $t-1$ and $t$, only candidates of these two stages are active. At the center of stage $t$ exactly the candidates of stage $t$ are active, so its guard gives $\sum_{a\in\Phi_t}x_{t,a}=z_t$, and summing over $t$ gives $|S|=K$; the first and the last stage have empty frontiers and targets $0$. Consider a vertex stage that processes $v$, with $\Phi_{t-1}=\Phi'\,\mathrm{In}(v)\,\Phi''$. Before the first replacement the coverage is $z_{t-1}$ by the center guard of stage $t-1$. Replacing the packet of an arc $a\in\Phi'$ changes the coverage by $x_{t,a}-x_{t-1,a}$, and the next guard has target $z_{t-1}$ again, so $x_{t,a}=x_{t-1,a}$; inductively this holds for all arcs of $\Phi'$. Replacing $\mathrm{In}(v)$ by $\mathrm{Out}(v)$ changes the coverage from $z_{t-1}$ to $z_{t-1}-\sum_{a\in\mathrm{In}(v)}x_{t-1,a}+\sum_{a\in\mathrm{Out}(v)}x_{t,a}$, and the guard demands $z_t$; since $z_t-z_{t-1}$ is the supply of $v$, the selected counts satisfy conservation at $v$. The arcs of $\Phi''$ are copied as those of $\Phi'$, now with target $z_t$. Identity stages copy every arc in the same way. Hence every arc $a$ has a single value $y_a\in\{0,\dots,B\}$ in all stages in which it lies on the frontier. Every arc of $\Gamma$ lies on some frontier, every vertex is processed once, and conservation holds at every vertex; so $y$ is a feasible flow.
\end{proof}

\begin{claim}\label{cl:realize}
Every feasible flow $y$ is realized by a selection that satisfies all guards: select any $y_a$ candidates of the packet of $a$ in every stage.
\end{claim}

\begin{proof}
At every center, the coverage is the total flow on the frontier, which is $z_t$. Each replacement in a vertex stage or an identity stage changes the coverage by the difference between new and old counts, which is zero for copied arcs and equals the supply of $v$ for the replacement of $\mathrm{In}(v)$ by $\mathrm{Out}(v)$; hence every guard sees its target.
\end{proof}

\begin{claim}\label{cl:probes}
If $S$ satisfies all guards, the probe of an arc $a$ is satisfied exactly when $y_a=0$. Every arc of $\Gamma$ has exactly one probe.
\end{claim}

\begin{proof}
In the identity stage after the vertex stage that created $a$, the coverage before the old packet of $a$ ends is $z_t$, by the preceding guard, and at the probe it is $z_t-x_{t-1,a}=z_t-y_a$, because the new packet has not started. Every arc is created exactly once, namely when its tail is processed.
\end{proof}

\begin{claim}\label{cl:proper}
The candidates form a proper family.
\end{claim}

\begin{proof}
All endpoints are distinct because all events have distinct coordinates. The starts of the candidates of stage $t$ lie between the centers of stages $t-1$ and $t$, and their ends between the centers of stages $t$ and $t+1$; within the stage, the starts appear in the order of $\Phi_t$, listed as $\Phi'\,\mathrm{Out}(v)\,\Phi''$ in a vertex stage, and the ends appear in the order of $\Phi_t$ as well, listed as the decomposition used by the next stage, which by the proof of \Cref{lem:frontier} is the same list; in both cases the candidates of one packet follow the order of $j$. Hence the starts and the ends of all candidates are both ordered by stage, then by frontier position, then by $j$.
\end{proof}

\begin{proof}[Proof of \Cref{lem:grid}]
By Claims~\ref{cl:guards}, \ref{cl:realize} and~\ref{cl:probes}, the selections satisfying all guards have size $K$, and their maximum value of $\Psi$ is $\Lambda g+\min_y|\{a:y_a>0\}|=\Lambda g+\OPT_{\mathrm{flow}}$, using \Cref{lem:flow}. A selection of size $K$ violating some guard has $\Psi(S)\le\Lambda(g-1)+e<\Lambda g$, since there are $e$ probes of weight $1$. Hence every maximizer satisfies all guards. Feasible flows exist, so $K$ is at most the number of candidates. Properness is \Cref{cl:proper}. The point weights are $1$ and $e+1$, hence positive integers bounded by the grid size. The grid has $|V(\Gamma)|$ vertices and $e\le2|V(\Gamma)|$ arcs, there are $2|V(\Gamma)|+1$ stages with at most $e$ frontier arcs each, and every packet has $B\le|V(\Gamma)|$ candidates; all quantities are polynomial in the size of the flow instance instance, because its coordinates are polynomially bounded.
\end{proof}

\subsection{Details of the reduction from exact coverage to demands}\label{app:tilt}

\paragraph*{Coordinates.}
Let $n_0$ be the number of intervals and $m_0$ the number of original and added points, before replacement; at most $2n_0-1$ points are added. For each such point $x$, let $t_x=T_x(K)$ be the number of replacement points created at $x$; then $t_x\le\Lambda K(n_0m_0+1)+\Lambda$. Let $\Delta=1+\max_xt_x$. Assign to the original and added points, from left to right, consecutive open segments of lengths $\Delta\omega_x$, with the segment for $x$ given by $(\Delta\varsigma_x,\Delta(\varsigma_x+\omega_x))$, where $\varsigma_x$ is the total mass of the points to its left. Move every endpoint to $\Delta$ times the total mass of the points to its left, and place the $t_x$ replacement points for $x$ at the integers $\Delta\varsigma_x+1,\dots,\Delta\varsigma_x+t_x$. These lie inside the segment for $x$ because $t_x<\Delta\le\Delta\omega_x$. Because an added point lies between any two consecutive endpoints, all endpoints remain distinct; every interval now has length $\Delta C$, all coordinates are distinct integers, no point is an endpoint, and each replacement point is covered by exactly the intervals that covered the point it replaces. Points whose demand exceeds the number of intervals covering them are never satisfied and are removed, which changes no number of satisfied points; afterwards every point $p$ satisfies $1\le q_p\le\min\{K,c_{\I}(p)\}$.

\paragraph*{Size.}
The number of replacement points is at most $m_0(\Lambda K(n_0m_0+1)+\Lambda)$, and all coordinates are at most $\Delta\,m_0(n_0m_0+1)$. Since $K\le n_0$ and $\Lambda$ is polynomially bounded, the construction has polynomial size and polynomially bounded numbers. The preceding reduction produces instances satisfying this weight bound, with $\Lambda=e+1$.

\paragraph*{Targets.}
For every integer reward threshold $R$, the \MaxEIMC{} instance has a selection of exactly $K$ intervals with reward at least $R$ if and only if the resulting \MaxIMC{} instance has a selection of at most $K$ intervals satisfying at least $Q=\Lambda CK-\Psi(\emptyset)+R$ points. This includes thresholds for which $Q\le0$ or $Q$ exceeds the number of replacement points.

\section{Trimming a solution that exceeds the budget}\label{app:trim}

Let $\eps>0$ and $K\ge1$ be such that $\eps K$ is an integer. Take the intervals $[i,i+K-\frac12]$ for $i=1,\dots,(1+\eps)K$ and points of demand $K$ at the integers $K,\dots,(1+\eps)K$. The point $j$ lies in exactly the $K$ intervals with $j-K+1\le i\le j$, so these $(1+\eps)K$ intervals together satisfy all $\eps K+1$ points, whereas any $K$ of them satisfy at most one point, because distinct points lie in distinct sets of $K$ intervals. Far to the right, add $K$ intervals that all contain $\eps K+1$ further points of demand $K$ and no other point. A set of at most $K$ intervals can satisfy points of only one of the two groups, so an optimal solution satisfies $\eps K+1$ points. The first $(1+\eps)K$ intervals satisfy as many points, but every $K$ of them satisfy at most one point.

\end{document}